\documentclass[pre,reprint,twocolumn,showpacs]{revtex4}

\usepackage{epsfig}
\usepackage[]{natbib}
\usepackage{color}
\usepackage{amsmath}
\usepackage[toc]{appendix}

\usepackage{amsfonts} 
\usepackage{color} 
\usepackage{hyperref}
\usepackage{graphicx,amssymb}
\usepackage{multirow}

\begin{document}
\title{
Infinite Ergodic Theory for Heterogeneous Diffusion Processes}

\author{N. Leibovich}
\author{E. Barkai}
\affiliation{Department of Physics, Institute of Nanotechnology and Advanced Materials, Bar-Ilan University, Ramat-Gan
5290002, Israel}


\begin{abstract}
We show the relation between processes which are modeled by a Langevin equation with multiplicative noise and infinite ergodic theory.  We concentrate on a spatially dependent diffusion coefficient that behaves as ${D(x)}\sim |x-\tilde{x}|^{2-2/\alpha}$ in the vicinity of a point $\tilde{x}$, where $\alpha$ can be either positive or negative.  We find that a nonnormalized state, also called an infinite density, describes statistical properties of the system. %In usual ergodic systems one finds that ensemble averages are computed using time averages and vise versa, where the ensemble averages are calculated from the invariant density. 
For processes under investigation, the time averages of a wide class of observables, are obtained using an ensemble average with respect to the nonnormalized density.  A Langevin equation which involves multiplicative noise may take different interpretation; It\^o, Stratonovich, or H\"anggi-Klimontovich, so the existence of an infinite density, and the density's shape, are both related to the considered interpretation and the structure of $D(x)$.       
\end{abstract}
\maketitle

\section{Introduction}
Consider a signal $x(t)$ which is modeled with a Langevin equation
\begin{equation}
\frac{dx}{dt}=\sqrt{D(x)}\eta(t),
\label{eq:LangevinMulti}
\end{equation}
where $D(x)$ is spatially dependent and $\eta(t)$ is a white noise with zero mean and $\langle \eta(t+t')\eta(t)\rangle =\delta(t')$. The initial condition  is $x(t)|_{t=0}=x_0$. This is a model for diffusion of a particle in an inhomogeneous system, where $x(t)$ is the trajectory of the particle.
%Model of multiplicative noise in an unidimensional environment, where the diffusion coefficient is space dependent, namely an heterogeneous diffusive process. 
Such spatially dependent diffusivities model many processes, where a partial list includes random walks in an inhomogeneous medium   \cite{lanccon2001drift,pevsek2016mathematical,regev2016isothermal}, chemical reactions \cite{Gardiner}, diffusion (in momentum space) in laser cooling processes \cite{CT92,bardou2002levy}, 
dissipative particle dynamics \cite{farago2016connection}, vortex-antivortex annihilations \cite{bray2000random}, studies of the stocks market in finances \cite{oksendal2013stochastic},  in  biophysics \cite{pieprzyk2016spatially,berezhkovskii2017communication,kuhn2011protein} e.g. measurements of proteins' diffusivity in mammalian cells  \cite{kuhn2011protein}, and modeling of $1/f^{\beta}$ noise %using  a non-linear differential equation 
\cite{Kaulakys}.

Importantly, care must be taken when dealing with multiplicative noise, since the Langevin equation may take different interpretations; It\^o, Stratonovich or H\"anggi-Klimontovich (isothermal) \cite{Gardiner,ito1944109,stratonovich1966new,hanggi1982nonlinear,klimontovich1990,kubo2012statistical}, see also Table~\ref{Tab1}. Generally, the interpretation of integration is related to the examined process and the nature of the noise %. On the different approaches and the choosing dilemma one can see 
  \cite{kubo2012statistical,Gardiner,wong1965,Volpe}. 
 The corresponding Fokker-Planck equation reads 
\begin{equation}
\frac{\partial P(x,t)}{\partial t} = \frac{1}{2}\frac{\partial}{\partial x}\left\lbrace D(x)^{1-\frac{A}{2}} \frac{\partial}{\partial x}\left[D(x)^{\frac{A}{2}} P(x,t)\right]\right\rbrace,
\label{eq:FP:General}
\end{equation}
with $A=0$  for H\"anggi-Klimontovich, $A=1$ for Stratonovich,  or $A=2$ for It\^o  interpretation. 
 Clearly, the solution of the Fokker-Planck equation, $P(x,t)$, depends on the behavior of $D(x)$ and the interpretation of Eq.~\eqref{eq:LangevinMulti}.

% Consider an ensemble of non-interacting one dimensional Brownian particles in the presence of a
% binding potential field $V(x)$. When the system reaches a thermal equilibrium the concentration is described by Boltzmann’s distribution $P_{\rm eq}(x) = \exp[- V(x)/k_BT ]/Z$ where $k_B T$ is the temperature (in units of energy) and $Z$ is the normalizing partition function. With this law we may calculate ensemble average of an observable ${\cal O}\left(x(t)\right)$; $\langle {\cal O}\rangle = \int {\rm d}x {\cal O}(x)P_{\rm eq}(x)$, where the observable ${\cal O}\left(x(t)\right)$ depends on the position of the particle $x(t)$. On the other hand we may construct the time average of the observable; ${\cal \overline{O}}_t = \int_0^t{\rm d}t' {\cal O}(t')/t$. The observable ${\cal O }$ may be the displacement $x$, $x^2$, sojourn time in a subspace, etc.  For an ergodic motion, the time and ensemble averages are identical in the limit of long measurement times $t\rightarrow\infty$ \cite{Gardiner,FellerBook, kubo2012statistical}, means $\lim_{t\rightarrow\infty}\overline{{\cal O}}_t=\langle {\cal O}\rangle$.

In the long time limit, a system may reach a steady state, namely $P(x,t)$  for long $t$ is time independent.
This solution is usually reached from most typical initial conditions, and the time-independent density is called the invariant density  \cite{InvariantMeasureBook}. For example when Brownian particles are confined in a finite domain, after a sufficiently long time their concentration becomes uniform (for reflecting boundary conditions) and thus time invariant. 
%Under some mathematical requirements the probability density function (PDF) of displacement $x$ at time $t$, $P(x,t)$, is an invariant density, also called an invariant measure, see  \cite{InvariantMeasureBook}. %In many times the invariant measure is given by $\lim_{t\rightarrow\infty}P(x,t)$.
Ergodic theory studies the properties of invariant densities.  For dynamical systems, Birkhoff's ergodic theorem states that if such an invariant density exists (i.e. it is normalizable) the ergodic assumption is fulfilled, namely the time-averaged observable converges to the average with respect to the normalized invariant measure \cite{InvariantMeasureBook}.
However, in some cases such an invariant state is nonnormalizable, and thus does not serve as a proper density. When the so-called infinite invariant density is found, a different type of ergodic framework emerges, and this is called infinite ergodic theory.  The mathematical concept of infinite densities was throughly investigated \cite{thaler1983transformations,thaler2006distributional}.

As was mentioned, the term ``infinite density'' refers to a  function which is nonnormalizable. Still, as we show below, this nonnormalizable state can describe statistical properties of the process.
At first glance this seems like a contradiction  since, as mentioned, a proper density is normalizable. Nevertheless, the infinite density captures some information on certain observables. 
%Namely, the infinite density describes some statistical properties of the process. %Infinite ergodic theory studies integrable and non-integrable observables with respect to the invariant density and examines the limit distribution of the time-averaged observable (see below). 
Our work is inspired by infinite ergodic theory which addresses deterministic paths, like the Pomeau-Manenville map \cite{korabel2009pesin,akimoto2010role}.
The concept was extended also to models
of laser-cooled atoms, L\'evy walks,  and non-equilibrium processes \cite{Kessler,rebenshtok2014infinite,aghion2018non}.
The statistical properties of observables that are integrable with respect to the invariant density are given by Aaronson-Darling-Kac theorem \cite{aaronson1997introduction,aaronson2005occupation}.

 In this paper, we demonstrate some features of infinite ergodic theory using a process which is modeled by a Langevin equation with multiplicative noise. 
In particular, we examine the  heterogeneous diffusion model with a power-law dependent diffusion coefficient in the vicinity of some point $\tilde{x}$, i.e. 
\begin{equation}
D(x)\propto |x-\tilde{x}|^{2-2/\alpha}.
\label{eq:DiffIntroduction}
\end{equation}
This, for example, is related to Richardson diffusion in turbulence %where $D(x)\propto x^{4/3}$
\cite{richardson1926atmospheric}, or generalized Lotka-Volterra equations modeling ecosystems %with $D(x)\propto x^{1/2}$ 
\cite{biroli2017marginally}. $\alpha=1$ is the ``normal'' case, where $D$ is simply a constant.
 It was shown that such processes yield anomalous diffusion, and the distribution of time-averaged mean-squared displacement was also considered, so it is known that
standard ergodic theory does not hold here \cite{cherstvy2013,Cherstvy,cherstvy2014nonergodicity,cherstvy2015ergodicity}.  The question is thus what is the proper ergodic theory for these anomalous processes? 
\. Here we show that the basic aspects mentioned above; %the double scaling solution, 
a limit state which is nonnormalizable, infinite ergodic theory, and the Aaronson-Darling-Kac theorem are applicable for this model as well. 

\begin{table}
\begin{tabular}{|p{2.6cm}c p{1.9cm}p{2.7cm}|}
\hline \hline
Model [Ref.] &  Form & $D(x)$ & Comments \\ \hline \hline
Vortex-Antivortex Annihilation  \cite{bray2000random} & S & $\propto 1/\ln  x $ & \\ \hline
1/f Noise \cite{Kaulakys} & I & $ \propto x^{2 \eta}$  & \\ \hline
\multirow{2}{*}{Nonlinear systems} &  & & \\ satisfying Einstein relation ~\cite{klimontovich1990} & HK & $\propto 1+ Bx$ \footnote{%The author examines nonlinear systems satisfied Einstein relation. %such as %Brownian particles with nonlinear friction.
This is given for Van der Pol oscillators. $B$ is a constant proportional to the temperature. } & 
\\ \hline
\multirow{2}{*}{Atmospheric} & & & $\partial_tP=$ \\
Diffusion \cite{richardson1926atmospheric} & & &  $ \epsilon \partial_x [x^{4/3}\partial_x P  ]$ \footnote{$\epsilon$ is a constant. %Here the Langevin equation is not specified.
}  \\ \hline
Ecosystems  \cite{biroli2017marginally} &  I & $ \propto x$ & \\ \hline
\multirow{2}{*}{Diffusion on }  & & & $\partial_t P= K x^{1-D}$  \\ a Fractal \cite{Procaccia1985analytical,havlin1987diffusion} 
& & & $  \partial_x [x^{-1-\theta +D }\partial_x P  ]$ \footnote{$K$ is a constant, $D$ is the fractal dimension. $\theta$ is related to the anomalous diffusion exponent.
%Langevin equation is not specified.
} \\
\hline \hline
\end{tabular}
\caption{Examples of models which have spatially dependent $D(x)$, with different interpretations of the Langevin equation \eqref{eq:LangevinMulti}; It\^o (I), Stratonovich (S), or H\"anggi-Klimantovich (HK). For some models, a form of a Langevin equation \eqref{eq:LangevinMulti} is not given, though an equation for the PDF $P(x,t)$ is provided, see \cite{richardson1926atmospheric,havlin1987diffusion,Procaccia1985analytical}. }
\label{Tab1}
\end{table}

%\section{Exact Solvable Example with positive $\alpha$}
\section{From Multiplicative Noise to Bessel Process}
\label{sec:PurePowerLaw}

In \cite{bray2000random}, Bray shows that a specific model of vortex-antivortex annihilation, which involves multiplicative noise, is closely related to the motion of a random walker in a central logarithmic potential, namely a Bessel process.  Here we extend this result and show that processes with $D(x)$ in the form of Eq.~\eqref{eq:Diff} (see below) are associated with the Bessel process as well.

Consider the Langevin equation \eqref{eq:LangevinMulti} with 
\begin{equation}
\sqrt{D(x)}=\sqrt{2D_0 \alpha^2} \left(\frac{x}{\ell}\right)^{1-\frac{1}{\alpha}}.
\label{eq:Diff}
\end{equation} 
The constant $D_0$ has units of $[\rm {cm}^{2}{\rm sec}^{-1}]$, and $\ell$ is some characteristic length scale. Generally the exponent $\alpha$ may be positive or negative. Currently, we concentrate on the case where $\alpha\geq 1$ and $x\in [0,\infty)$, so the growth condition is fulfilled (see \cite{Gardiner} and App.~\ref{App:Growth}), thus we ensure stability of the paths.  We also require $(1-A)(1-\alpha)<1$ for a reason that we will clarify soon.  Initially, all particles are located in $x(t)|_{t=0}=x_0$. At $x=0$ we use a reflecting boundary condition. 
%Other cases of $\alpha$ are studied below. 
% Note that if a particle has begun at zero it stuck there forever. Furthermore, if starting at any other point, when approaching to zero the particle becomes slower and slower, since $D(x)\rightarrow 0$ when $x\rightarrow 0$. %so actually the particle will never reach zero. Therefore, zero is considered an unstable fixed point. 
The specific choice in Eq.~\eqref{eq:Diff} allows an exact treatment of the problem for any time $t$. Later we consider a more general form of the diffusion field.

There is a known mapping between It\'o and H\"anggi-Klimontovich forms  of Eq.~\eqref{eq:FP:General} to Stratonovich interpretation (see e.g. \cite{klimontovich1990} and App.~\ref{Sec:Appendix1}). The Fokker-Planck equation \eqref{eq:FP:General} is rewritten
\begin{eqnarray}
\frac{\partial P(x,t)}{\partial t} 
&=& \frac{1}{2}\frac{\partial}{\partial x}\left[\sqrt{D(x)} \frac{\partial}{\partial x} \sqrt{D(x)}P(x,t)\right] \nonumber \\
&-&\frac{1-A}{2}\frac{\partial}{\partial x}\left[\sqrt{D(x)}\frac{\partial \sqrt{D(x)}}{\partial x}P(x,t)\right].
\label{eq:FP_Stratonovich}
\end{eqnarray}
Thus,  the Stratonovich interpretation of a Langevin equation with an additional effective drift term (i.e. the second term on the right-hand side in Eq.~\eqref{eq:FP_Stratonovich}) is equivalent to the Langevin equation \eqref{eq:LangevinMulti} with the H\"anggi-Klimontovich ($A=0$) or It\^o ($A=2$) forms. Its corresponding Langevin equation is
\begin{equation}
\frac{dx}{dt}=\sqrt{D(x)} \eta(t) + \frac{1-A}{2}\sqrt{D(x)}\frac{d\sqrt{D(x)}}{dx},
\label{eq:LangevinMultiTranformed}
\end{equation}
which is now interpreted via the Stratonovich approach.

Now we define the transformation \cite{ cherstvy2013,bray2000random}
\begin{equation}
y(x)\equiv\int_0^x\frac{{\rm d}x}{\sqrt{D(x)}}=\frac{\ell^{1-\frac{1}{\alpha}}}{\sqrt{2D_0}}x^{\frac{1}{\alpha}}
\label{eq:Transformation}
\end{equation}
where $y\in [0,\infty)$ and $y_0 \equiv y(x_0)$. 
The above transformation may be used only when interpreting the noise as continuous, namely in the Stratonovich form (i.e. following Wong-Zakai theorem \cite{wong1965}). 
Therefore we obtain that Eq.~\eqref{eq:LangevinMultiTranformed} is mapped to 
\begin{equation}
\dot{y}=\eta(t)+\frac{1-A}{2}\cdot\frac{d\sqrt{D(y)}}{dy}\frac{1}{\sqrt{D(y)}},
\label{eq:y}
\end{equation}
then, using Eqs.~\eqref{eq:Diff} and \eqref{eq:Transformation} we find
\begin{equation}
\dot{y}=\eta(t)-\frac{U_0/2}{y},
\label{eq:Langevin_y}
\end{equation}
where $U_0=(1-A)(1-\alpha)$. The variable $y$ describes the position of a Brownian particle in a logarithmic potential so the additional effective force is given by $F(y)=-\frac{1}{2}U_0/y=-\frac{1}{2}U_0\partial_y \ln y $. Note that the potential can be repulsive or attractive. Eq.~\eqref{eq:Langevin_y} is the Bessel process which is related to the diffusion of particles in high dimension, where $y$ is the radial displacement, and $U_0$ is associated with the dimension \cite{bray2000random,martin2011first}.
The probability density function (PDF) of $y$ in time $t$, with the initial condition $P(y,t)|_{t=0}=\delta(y-y_0)$, and the reflecting boundary condition, i.e.  $\partial_y P(y,t)|_{y=0}=0$,   is  
\begin{equation}
P(y,t;y_0,0)=e^{-\frac{y^2+y_0^2}{2t}}y_0^{\frac{1}{2}+\frac{U_0}{2}}y^{\frac{1}{2}-\frac{U_0}{2}}I_{-\frac{1}{2}-\frac{U_0}{2}}\left(\frac{y_0y}{t}\right)\frac{1}{t},
\end{equation}
where $I_{\nu}(z)$ refers to the modified Bessel function of the first kind of order $\nu$ \cite{bray2000random,martin2011first}. This PDF is normalized when $U_0<1$, so here $(1-A)(1-\alpha)<1$ as mentioned. %(for stronger repulsive potential the particles escape too fast to infinity).
% In the limit of small $y_0$ ($y_0 \ll y$) we obtain
% \begin{equation}
% P(y,t)\approx {\cal N} t^{-\frac{1}{2}+\frac{U_0}{2}} y^{-U_0}e^{-\frac{y^2}{2t}}
% \end{equation}
% \cite{bray2000random,martin2011first}.  Back to $P(x,t)$ we find
% \begin{equation}
% P(x,t)\approx {\cal N}x^{A\left(-1+\frac{1}{\alpha}\right)}\exp\left(-\frac{x^{\frac{2}{\alpha}}}{4D_0 t} \right) t^{-\frac{\alpha}{2}+\frac{A}{2}(\alpha-1)}
% \label{eq:TimeDependentItoHanggi}
% \end{equation}
% where ${\cal N}$ is the normalization constant of $P(x,t)$. 
Back to $P(x,t)$ using Eq.~\eqref{eq:Transformation} we find that
\begin{eqnarray}
&&P(x,t)= {\cal N} \exp\left[-\frac{(x^{\frac{2}{\alpha}}+x_0^{\frac{2}{\alpha}})\ell^{2-\frac{2}{\alpha}}}{4D_0 t}\right] 
\label{eq:TimeDependentItoHanggi}
\\ \nonumber
&&x_0^{\frac{1}{2\alpha}\left(1+U_0\right)}x^{\frac{1}{2 \alpha}\left(3-U_0-2\alpha \right)}I_{-\frac{1}{2}-\frac{U_0}{2}}\left(\frac{x_0 ^{\frac{1}{\alpha}}x^{\frac{1}{\alpha}}\ell^{2-\frac{2}{\alpha}}}{2D_0 t}\right)\frac{1}{t}
\end{eqnarray}
where ${\cal N}=\ell^{2-\frac{2}{\alpha}
}/[2D_0\alpha]$ is the normalization constant. It is easy to verify that Eq.~\eqref{eq:TimeDependentItoHanggi} is the normalized solution of Eq.~\eqref{eq:FP:General},
with the initial condition $P(x,0)= \delta(x-x_0)$. We note that the following results are also valid for sufficiently long time for other initial concentrations which are inherently narrow, e.g. Gaussian distribution centered in $x_0$.

{\bf Comment:} Mathematically, the above solution, Eq.~\eqref{eq:TimeDependentItoHanggi}, exists when $ U_0 \equiv (1-A)(1-\alpha)<1$. For stronger effective potential, when $U_0\geq 1$, the particles fall to the origin, thus the only solution is when zero serves as an absorbing point, see discussion in \cite{bray2000random}. A regularization of the diffusion at the vicinity of the origin settles the problem with $U_0\geq 1$, as commented in \cite{bray2000random} and we show below in Sec.~\ref{sec:Regularized}. 

We will soon relax the conditions made in this section. The requirements $U_0<1$ with $\alpha\geq 1$, limit the range of $\alpha$ for It\^o interpretation, so here $1\leq\alpha<2$ for $A=2$. For H\"anggi-Klimontovich ($A=0$) and Stratonovich ($A=1$) interpretations we use $\alpha \geq 1$. Moreover we note that our results are valid for finite $\alpha$ only, where  essentially different results are obtained in the limit $\alpha\rightarrow\infty$. This case, where $D(x)\propto x^2$, is not of the scope of this paper and is excluded.

\subsection{Infinite Density}
To gain insight on the long-time limit of the solution we write Eq.~\eqref{eq:FP:General} as $\partial_t P=-\partial_x J$, where 
\begin{equation}
J\equiv -\frac{1}{2}D(x)^{1-\frac{A}{2}}\partial_x\left[D(x)^{\frac{A}{2}} P(x,t)\right].
\end{equation}
In many circumstances, when setting $J=0$ the steady-state solution $P(x,t)=P_{ss}(x)$, which is an invariant density, is obtained. In our case, there is no steady state in the usual sense, but still we search for a solution ${\cal I}_{\infty}(x)$ 
that satisfies 
\begin{equation}
D(x)^{1-\frac{A}{2}}\partial_x\left[D(x)^{\frac{A}{2}} {\cal I}_{\infty}(x)\right]=0,
\label{eq:CurrentDef}
\end{equation}
which is an infinite density.
Here, the solution of zero current, $J=0$, obtained from Eq.~\eqref{eq:CurrentDef}, is 
 \begin{equation}
{\cal I}_{\infty}(x)= C D(x)^{-\frac{A}{2}} = C \frac{1}{(2D_0 \alpha ^2)^{A/2}}\left(\frac{x}{\ell}\right)^{A \left(-1+\frac{1}{\alpha}\right)}.
\label{eqLSteadyState}
\end{equation}
While solving $\partial _t P=0$ one finds another solution which diverges when $x$ goes to infinity, hence cannot capture a physical sense, thus the only solution is given when $J=0$.
However, the solution Eq.~\eqref{eqLSteadyState}, is not normalizable, hence as a stand alone solution it is not valid. Therefore, the constant $C$ is not related to the normalization in the usual way. 
Note that since $U_0<1$ and $\alpha\geq 1$ [equivalent to $-1<A(-1+1/\alpha)\leq 0$] the divergence in the spatial integral $\int_0^{\infty} {\rm d}x{\cal I}_\infty(x)$ is caused by the large $x$ behavior of ${\cal I}_{\infty}(x)$. Importantly, note that there is a relation between the nonnormalizable zero-current solution Eq.~\eqref{eqLSteadyState} and the time-dependent distribution Eq.~\eqref{eq:TimeDependentItoHanggi} via
\begin{eqnarray}
&&\lim_{t \rightarrow\infty }P(x,t)t^{\frac{\alpha}{2}-\frac{A}{2}(\alpha-1)} = 
{\cal I}_{\infty}(x)   \nonumber \\ 
&& = C \frac{1}{({2D_0 \alpha ^2 })^{A/2}}\left(\frac{x}{\ell}\right)^{A \left(-1+\frac{1}{\alpha}\right)}, 
\label{eq:TimeDependentItoHanggiLimit}
\end{eqnarray}
where  $C={2^{1-\alpha(1-A)-\frac{A}{2}}D_0^{-\frac{\alpha(1-A)}{2}}\ell^{-U_0}|\alpha|^{A-1}}/{\Gamma\left[\frac{1-U_0}{2}\right]}$. This solution is called  an infinite density in the sense that it is nonnormalizable.
From Eq.~\eqref{eq:TimeDependentItoHanggiLimit} it is easy to understand why ${\cal I}_\infty(x)$ is not normalized. On the left-hand side we have $P(x,t)$ times a prefactor that increases with time. Since the area under $P(x,t)$ is unity, but $t^{(\alpha-A\alpha+1)/2}\rightarrow\infty$, clearly the
integral over ${\cal I}_{\infty}$ must blow up. 
More surprising is that this nonnormalized state captures some of the 
physical properties of the process as is shown below. 

\subsection{Infinite Ergodic Theory}
\label{sec:PurePowerLawInfinite Ergodic Theory}
Consider an observable ${\cal O}[x(t)]$, which depends on the realization $x(t)$. Assume that the observable ${\cal O}[x(t)]$ 
fulfills the following requirement
\begin{equation}
 \int_0^{\infty}{\rm d}x {\cal O}[x] {\cal I}_{\infty}(x)<\infty,
 \label{eq:AverageRspectToInfinite}
\end{equation} 
namely the observable is integrable with respect to ${\cal I}_{\infty}(x)$. The time average of ${\cal O}[x(t)]$  is defined as
\begin{equation}
\overline{ {\cal O}} _{t}\equiv \frac{1}{t}\int_0^{t}{\rm d}t' {\cal O}[x(t')],
\end{equation}
and the ensemble average reads
\begin{equation}
\langle {\cal O}_t\rangle \equiv \int_0^{\infty}{\rm d}x {\cal O}[x] P(x,t).
\end{equation} 
Generally both $\overline{ {\cal O}} _{t}$ and $\langle {\cal O}\rangle$ are time dependent. In the long time limit, using Eq.~\eqref{eq:TimeDependentItoHanggiLimit}, we obtain
\begin{equation}
\langle {\cal O}_t\rangle \stackrel{t\rightarrow\infty}{\approx} t^{\beta-1} \int_0^{\infty}{\rm d}x {\cal O}[x] {\cal I}_{\infty}(x),
\label{eq:Ensemble}
\end{equation}
with
\begin{equation}
\beta=1-\frac{\alpha}{2}-\frac{A}{2}(1-\alpha)=\frac{1}{2}+\frac{U_0}{2}.
\label{eq:beta}
\end{equation}
Now consider the ensemble average of the time average 
\begin{eqnarray}
\langle\overline{\cal O}_{t}\rangle &\equiv& \int_0^{\infty} {\rm d}x P(x,t) \frac{1}{t}\int_0^t{\rm d}t' {\cal O}[x(t')] \\
\nonumber
&\equiv& \frac{1}{t}  \int_0^t{\rm d}t'\int_0^{\infty} {\rm d}x  {\cal O}[x]P(x,t').
\end{eqnarray}
Therefore we find
\begin{eqnarray}
\langle\overline{\cal O}_{t}\rangle &\equiv& \frac{1}{t}  \int_0^t{\rm d}t'\int_0^{\infty} {\rm d}x  {\cal O}[x]P(x,t') 
\label{eq:TimeEmsemble}
\\
\nonumber
&\stackrel{t\rightarrow\infty}{\approx}& \frac{1}{t}  \int_0^t{\rm d}t' t'^{\beta-1} \int_0^{\infty} {\rm d}x  {\cal O}[x]{\cal I}_{\infty}(x) \\
\nonumber
&=& \frac{t^{\beta-1}}{\beta}  \int_0^{\infty} {\rm d}x   {\cal O}[x]{\cal I}_{\infty}(x),
\end{eqnarray}
where the prefactor $1/\beta$ [see Eq.~\eqref{eq:beta}] comes from the time integration. Hence, using Eqs.~\eqref{eq:Ensemble} and \eqref{eq:TimeEmsemble}, we conclude that
\begin{equation}
\lim_{t{\rightarrow\infty}}\frac{ \beta \langle \overline{{\cal O}} _{t}\rangle }{\langle {\cal O}_t\rangle }=1.
\label{eq:ZetaGoToOne}
\end{equation} 
Thus, time and ensemble averages are related, and the limit $\beta\rightarrow 1$ corresponds to the standard ergodic theroy. 
 
Furthermore, $\overline{\cal O}_{t}$ is a stochastic variable determined by the trajectory of $x(t)$ thus we define the random variable 
\begin{equation}
\xi\equiv \lim_{t{\rightarrow\infty}}\frac{ \beta \overline{ {\cal O}} _{t}}{\langle {\cal O}_t\rangle}.
\label{eq:ZetaDefine}
\end{equation}
In the following we examine the PDF of $\xi$ (clearly with $\langle \xi \rangle =1$), where $0<\beta<1$ (i.e. $-1<U_0<1$, weak potential).

For example let us consider the observable ${\cal O}[x(t)]=\theta(0.4<x(t)<0.6)$ which is a pulse function. Thus ${\cal O}[x(t)]$ alternates between ${\cal O}[x(t)]=1$ when $x(t)\in (0.4,0.6)$ and ${\cal O}[x(t)]=0$ otherwise. 
The time integration of the pulse function is the occupation time in the domain, so $\overline{\cal O}_t$ is the time spent by the process in the interval $(0.4,0.6)$ divided by the measurement time.
Using \cite{bray2000random,redner2001guide,martin2011first} we deduce that the sojourn times (i.e. first passage time) PDF when ${\cal O}[x(t)]=0$ (i.e. outside the interval) follows
\begin{equation}
\psi(\tau)\sim \tau^{-1-\beta},
\label{eq:Psi}
\end{equation}
in the long time limit (see \cite{bray2000random,redner2001guide,martin2011first} and App.~\ref{AppC}). Here the average sojourn time of the particle beyond the observation domain [outside the interval $(0.4,0.6)$] diverges   $\langle \tau\rangle =\infty$ since $0<\beta<1$. The number of times $x(t)$ re-enter the interval under observation until time $t$ is $n(t)$, $\overline{\cal O}_t \propto n $ and since $\langle \xi \rangle =1$ (as mentioned) we have 
\begin{equation}
\xi\equiv \beta\overline{\cal O}_t/\langle {\cal O} \rangle \sim n(t) /\langle n \rangle.
\end{equation}
This equation means that the distribution of the normalized time-average is the same as the distribution of the number of renewals. From the renewal processes studies we know that the number of renewals up to time $t$ divided with its mean (i.e. the variable $\xi$) is given by Mittag-Leffler distribution of order of $\beta$, ${\cal M}_{\beta}(\xi)$, see e.g. \cite{Godreche,metzler2014anomalous}. Therefore the distribution of $\xi$ is expected to follow the  Mittag-Leffler distribution as well, i.e.
\begin{equation}
{\rm P}(\xi)={\cal M}_{\beta}(\xi)\equiv \frac{\Gamma ^{\frac{1}{\beta}}(1+\beta)}{\beta \xi ^{1+\frac{1}{\beta}}} L_{\beta}\left[\frac{\Gamma ^{\frac{1}{\beta}}(1+\beta)}{\beta \xi ^{\frac{1}{\beta}}}\right],
\label{eq:zeta}
\end{equation}
where $L_{\beta}(z)$ is the one-sided L\'evy density of order $\beta$, which is defined through the following inverse Laplace transform from $s $ to $z$; $ L_{\beta}(z)\equiv {\cal L}^{-1}\left[\exp(-s^\beta)\right]$, see App.~\ref{AppD}.  
The above argument, Eq.\eqref{eq:zeta},  also applies to any observable which fulfills Eq.~\eqref{eq:AverageRspectToInfinite}, namely where it is integrable
with respect to an infinite measure of a system  \cite{aaronson2005occupation}. This result is in the spirit of the Aaronson-Darling-Kac theorem
 usually applied in the context of deterministic setting
\cite{aaronson1997introduction}.

For some intuition of the results consider a free Brownian particle with realization $y(t)$. There, the sojourn times $\tau$ of the trajectory $y(t)$ outside a given finite interval are distributed with $\psi(\tau)\sim \tau^{-3/2}$. The transformation $y(x)$ given in Eq.~\eqref{eq:y} is stretching or compressing the space in such a way that the temporal properties such as the return times behave similarly for $y$ and $x$. Therefore, the sojourn times of the realization $x(t)$ with Stratonovich interpretation (which is mapped into a free Brownian particle) 
outside a finite interval in $x$ space, 
%share a similar temporal behavior of the free Brownian realizations $y$, namely the sojourn times of $x(t)$ outside a finite interval 
is $\psi(\tau)\sim \tau^{-3/2}$, namely $\beta=1/2$.  For It\^o and H\"anggi-Klimontovich interpretations  the results, Eqs.~\eqref{eq:ZetaGoToOne}
\eqref{eq:zeta} with \eqref{eq:beta}, are similar to the ones found in diffusion in Logarithmic potential \cite{aghion2018non}. Roughly speaking, now with the mapping to Bessel processes at our hand, we can apply these general results, to the case under study here:
diffusion in inhomogeneous medium.

\subsection{Simulation Results}

\begin{figure}
\includegraphics[trim= 90 200 80 240, width=\columnwidth]{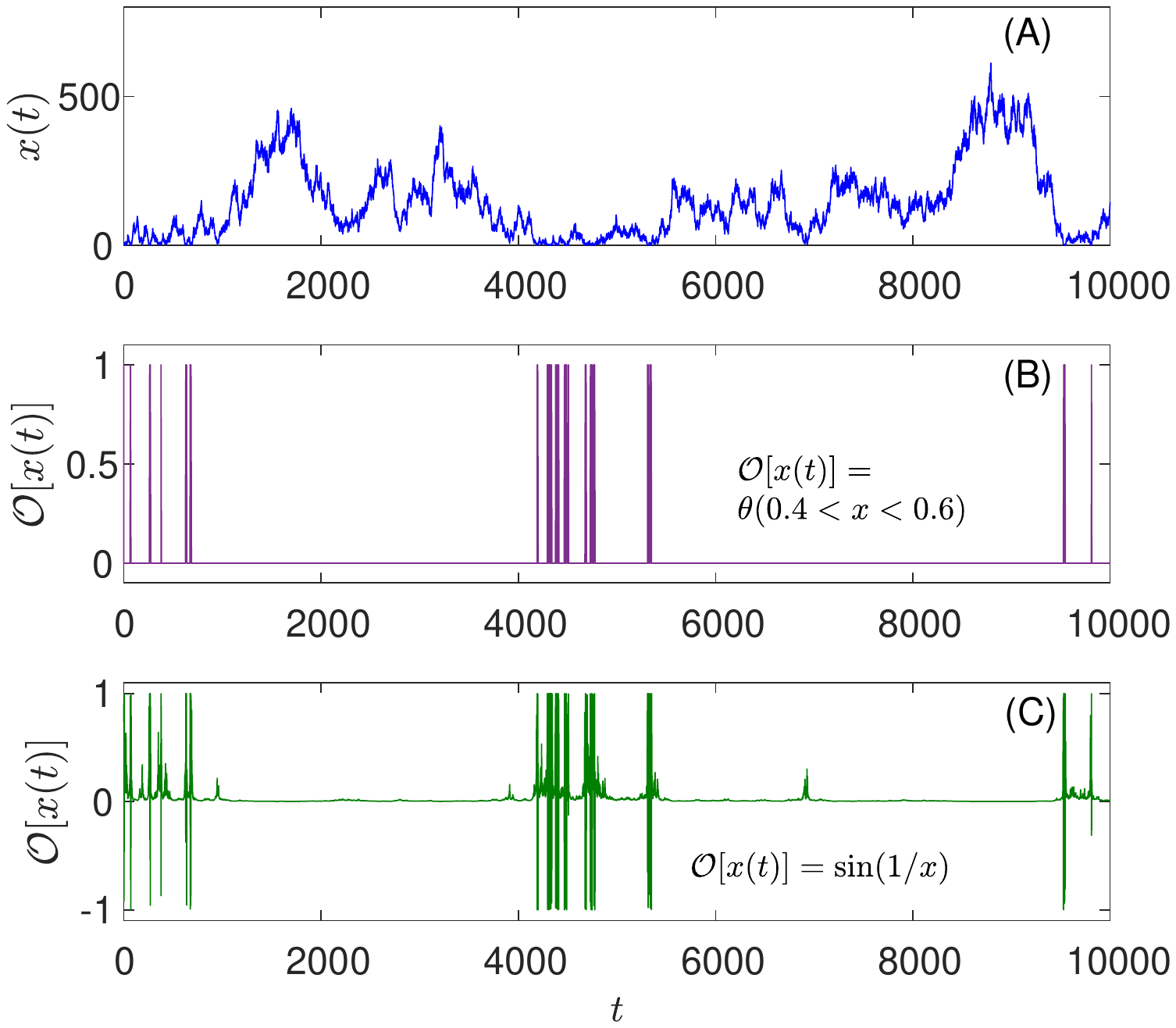}
\caption{ A trajectory $x(t)$ [panel (A)] and its corresponding observable ${\cal O}[x(t)]=\theta(0.4<x(t)<0.6)$ [panel (B)] and ${\cal O}[x(t)]=\sin\left[1/x(t)\right]$ [panel (C)[. The signal is generated from Langevin equation \eqref{eq:LangevinMulti} with \eqref{eq:Diff} and Stratonovich interpretation. 
Here we use $\alpha=3/2$ and the measurement time is $10^4$. The long sojourn times of ${\cal O}[x(t)]$ close to zero are visible.       }
\label{fig:SingleRealization}
\end{figure}

In Fig.~\ref{fig:SingleRealization} we demonstrate the long sojourn times close to zero. We generate a trajectory $x(t)$ from the Langevin equation \eqref{eq:LangevinMulti} with \eqref{eq:Diff} and Stratonovich interpretation. We use $\alpha=3/2$ and the measurement time is $10^4$. In all simulations in this paper we use $D_0=1/2$ and $\ell=1$.  $x(t)$ is given in the panel (A). In panel (B) we present the observable ${\cal O}[x(t)]=\theta(0.4<x(t)<0.6)$ where $x(t)$ is the same realization given in (A). 
%The long sojourn times outside the interval $(0.4,0.6)$ are visible. 
In panel (C) we show the mathematical observable ${\cal O}[x(t)]=\sin\left[1/x(t)\right]$, which is chosen to demonstrate the fact that the choice of a specific observable is not important. Both observables [in panels (B) and (C)] are integrable with respect to the infinite density. Therefore, they share a similar property; the observables have long sojourn times close to zero, thus  Eq.~\eqref{eq:zeta} with $\beta=1/2$ in agreement with Eq.~\eqref{eq:beta} is valid, see Fig.~\ref{fig:SingleRealization}.

\begin{figure}
\includegraphics[trim= 90 200 90 230, width=\columnwidth]{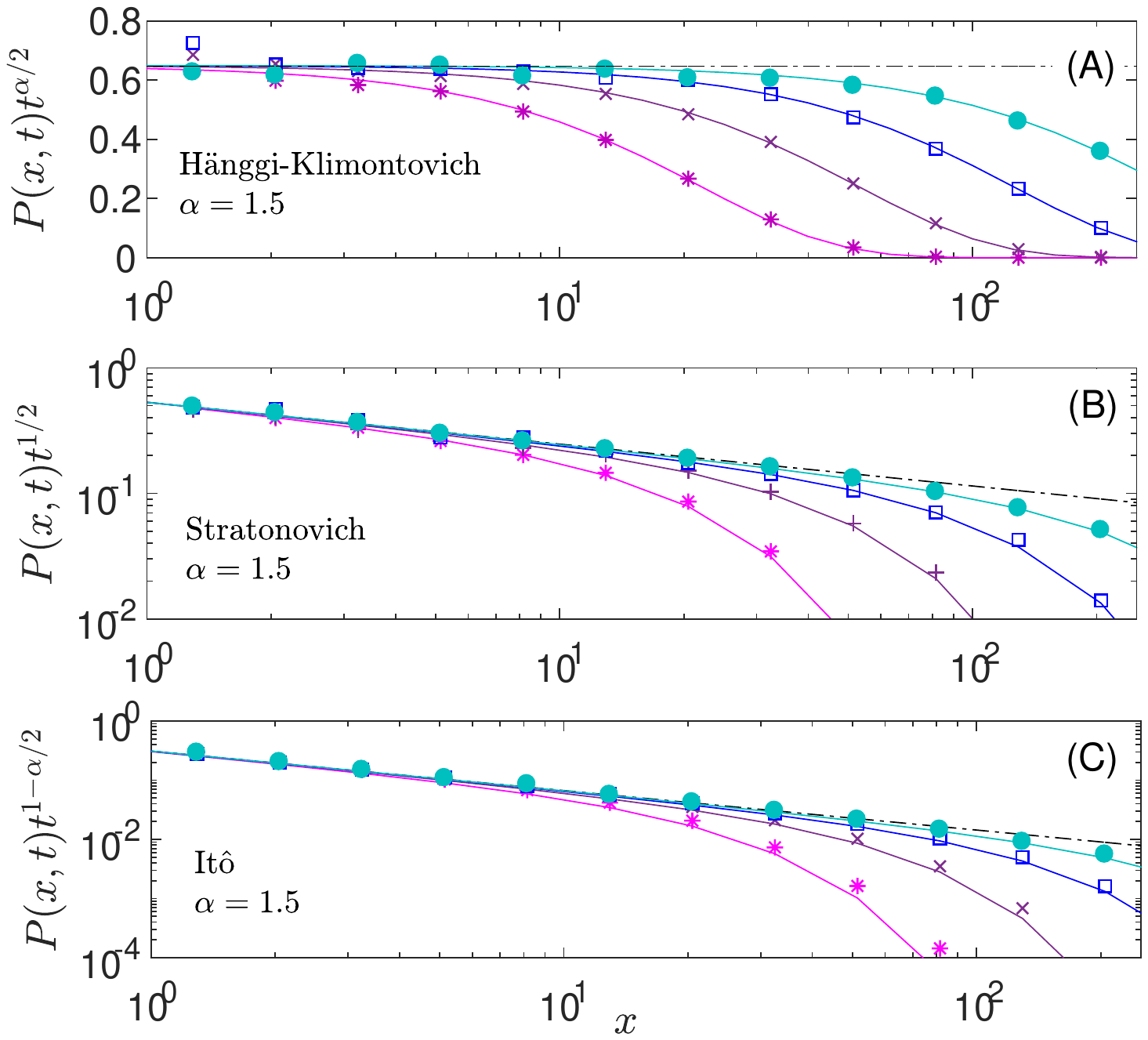}
\caption{ The scaled PDF $t^{1-\beta}P(x,t)$ for different times with $\sqrt{D(x)}=\alpha x^{1-1/\alpha}$ where $\alpha=3/2$ and $x>0$. Panel (A) presents the results for H\"anggi-Klimontovich interpretation, panel (B) for Stratonovich interpretation and panel (C) shows the results for the It\^o interpretation. Here we present the simulation results for $t=31$ (pink stars), $t=100$ (purple crosses), $t=316$ (blue squares) and $t=1000$ (cyan full circles).  The number of particles is $10^5$. Note that the upper panel is presented in semi-log scale while the other panels are given in double-log scale. An agreement between simulation results (symbols), the analytical prediction Eq.~\eqref{eq:TimeDependentItoHanggi} (solid lines), and the limit behavior Eq.~\eqref{eq:TimeDependentItoHanggiLimit} (dashed line) is shown. 
In the long limit the nonnormalized state ${\cal I}_{\infty}(x)$ is approached,
even though $P(x,t)$ is normalized for any finite time. 
}
\label{fig:P_x_t_ItoHanggi}
\end{figure}

% Consider the Langevin equation \eqref{eq:LangevinMulti} where $\sqrt{D(x)}$ is given in Eq.~\eqref{eq:Diff}. 
In Fig.~\ref{fig:P_x_t_ItoHanggi} we present the simulation results of $P(x,t)$  for processes $\sqrt{D(x)}=\alpha x^{1-1/\alpha}$ where $\alpha=3/2$. Panel (A) presents the results for H\"anggi-Klimontovich interpretation $(A=0)$, panel (B) for Stratonovich ($A=1$), and (C) shows results for It\^o interpretation $(A=2)$.    The agreement between the simulation results (symbols), the time-dependent solution Eq.~\eqref{eq:TimeDependentItoHanggi} (solid lines) and the limit distribution Eq.~\eqref{eq:TimeDependentItoHanggiLimit} (dashed lines) is visible.
 These simulations clearly demonstrate that the nonnormalized state
is measurable. Of-course for finite times we see deviations, however as we increase the measurement time, the 
nonnormalized state is approached.

\begin{figure}
\includegraphics[trim= 110 260 110 280, width=\columnwidth]{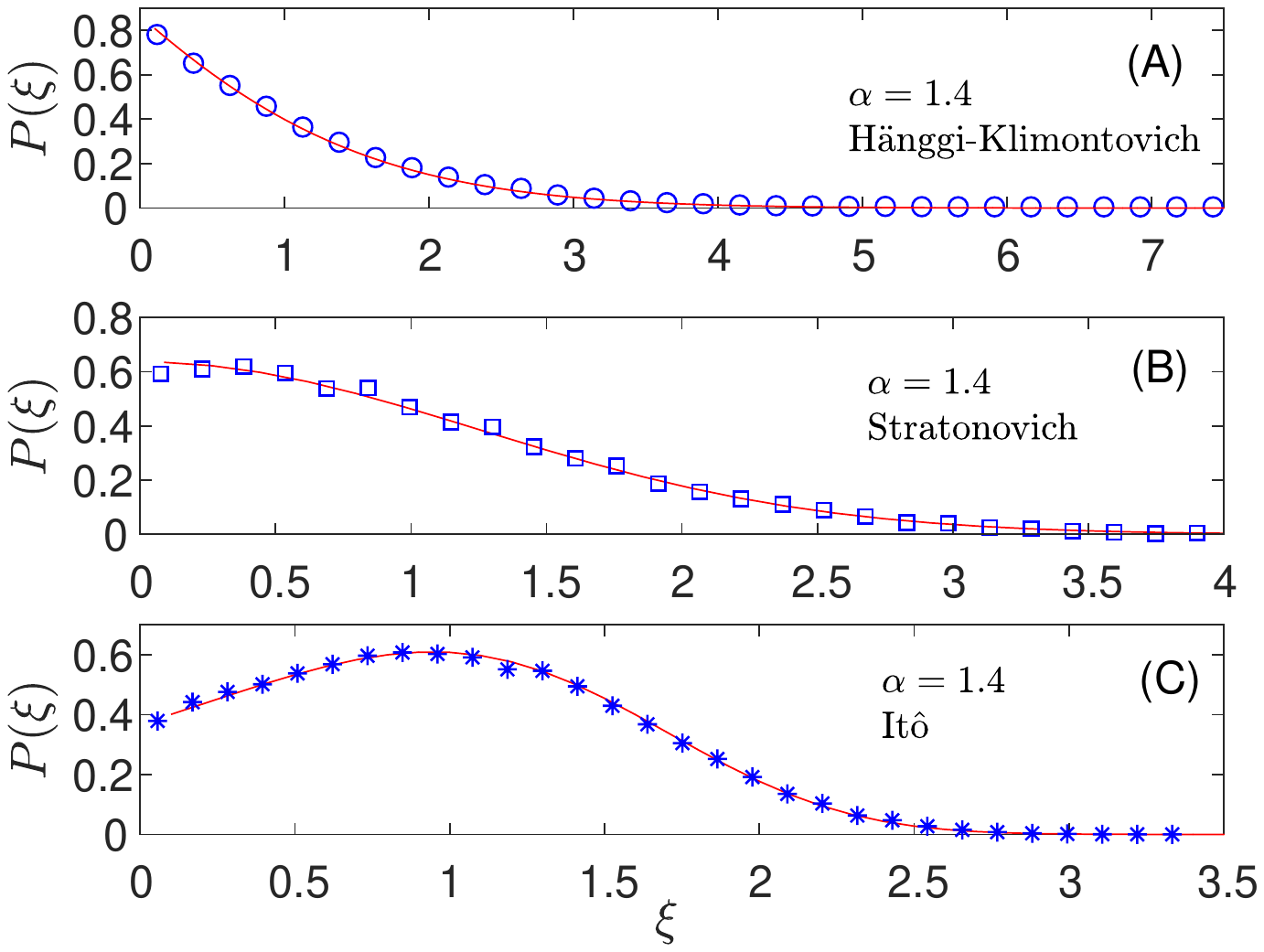}
\caption{ The distribution of the random variable $\xi$ defined in Eq.~\eqref{eq:ZetaDefine} with $
{\cal O}[x(t)]=\theta(0.4<x(t)<0.6)
 $ and $D(x)$ is given in Eq.~\eqref{eq:Diff} with $\alpha=1.4$. The simulation results are presented in blue circles [panel (A), H\"anggi-Klimontovich], blue rectangles [panel (B), Stratonovich] and blue stars [panel (C), It\^o]. For the simulation we use $10^5$ particles, $\alpha=1.4$ and $t=10^3$. The red curves represent the analytic predictions; Mittag-Leffler of order $\beta=0.3$  (A), 0.5 (B), and 0.7 (C), see Eq.~\eqref{eq:zeta}  }
\label{fig:P_xi_ItoHanggiAlpha1_4}
\end{figure}

As explained above, for an observable   ${\cal O}[x(t)]$ which is integrable with respect to ${\cal I}_{\infty}(x)$ infinite ergodic theory holds. For the illustration we choose 
$
{\cal O}[x(t)]=\theta(0.4<x(t)<0.6)
 $
 and define the random variable
$\xi$ using Eq.~\eqref{eq:ZetaDefine}. Then, in the long time limit, ${\rm P}(\xi)$ follows the Mittag-Leffler distribution of order $\beta$.
In Fig.~\ref{fig:P_xi_ItoHanggiAlpha1_4} we present the simulation results (with symbols) for the PDF of $\xi$ where $\alpha=1.4$, $t=10^3$ and $10^5$ particles. Panel (A) presents results for H\"anggi-Klimontovich (Mittag-Leffler function of order 0.3), panel (B) presents the results for Stratonovich (Mittag-Leffler of order 0.5) and (C) for It\^o interpretation (Mittag-Leffler or order 0.7).
Here we demonstrate that the statistics of $\xi$ depends on the stochastic interpretation of the Langevin equation \eqref{eq:LangevinMulti}.

% Furthermore, following Aaronson-Darling-Kac theorem, $\zeta$,
% which is a random variable, is generated from the Mittag-Leffler distribution of order $\beta$. The intuition of this result is the following. Above we find an infinite density $P_{\infty}(x)$ which is given in the infinite time limit of $P(x,t)$. It means that a particle is getting localized in a subspace of large $x$, means that while the measurement time increases the particle stays longer and longer in the $|x|>x^*$ subspace where $x^*$ is an arbitrary positive finite number. Furthermore, from the power-law dependent of the distribution, one can obtains that the sojourn times in the subspace, $\{\tau\}$, are distributed with a common PDF $\psi(\tau)\sim \tau^{-1-\beta}$ where $0<\beta<1$ \cite{bray2000random,redner2001guide}. Now, consider an observable ${\cal O}[x(t)]$  which is integrable with respect to the infinite density.  In that case the observable is dominant by these long sojourn times of $x(t)$ in the subspace $|x|>x^*$. Therefore the term $\int_0^t{\rm d}t{\cal O}[x(t)]$ is controlled by the number of times $x(t)$ re-enter the subspace and its average in this region. From the renewal processes we know that the number of renewals up to time $t$ is given by Mittag-Leffler distribution in the order of $\beta$, see e.g. \cite{Godreche,InvariantMeasureBook,Margolin04}, therefore the distribution of $\zeta$ is expected to follows Mittag-Leffler distribution as well. 

\subsection{Other Structures of $D(x)$ }
\label{sec:Regularized}

% In some cases, the diffusion is only asymptotically spatially power-law dependent. 
%  The additional effective force in Eq.~\eqref{eq:y}, which is proportional to $ {d_y \sqrt{D(y)}}/{\sqrt{D(y)}}$, is determined by the properties of $\sqrt{D(y)}$.
%  We assume that in some region, e.g. next to some $\tilde{y}$, one finds $\sqrt{D(y)}\sim |y-\tilde{y}|^{p}$ and $d_y \sqrt{D(y)} \sim |y-\tilde{y}|^{p-1}$ where $p \neq 0$, therefore a logarithmic like potential is obtained in that region, means that the effective force goes like $ |y-\tilde{y}|^{-1}$ when $y \rightarrow \tilde{y}$. 
As was mentioned in the Introduction, we study processes where, in the vicinity of some point $\tilde{x}$, the  diffusion coefficient is Eq.~\eqref{eq:DiffIntroduction}. 
%However, it does not mean that the spatially dependence of $D(x)$ is purely a power-law.
% Here we demand that the growth condition is fulfilled (see App.~\ref{App:Growth} and \cite{Gardiner}). This apply some constrains on the values of $\alpha$ and the whether $x(t)$ is a bounded process or not.  
In the previous subsections we considered a specific form of $D(x)$, Eq.~\eqref{eq:Diff}, which allowed us to obtain exact results for any time $t$. Our aim now is to show that
the features such as the infinite density are generally valid, in particular for processes with $D(x)\neq 0$ on $x=0$.   
For example consider a process with %a regular constant diffusion which characterized by some length scale;
\begin{equation}
\sqrt{D(x)}=\sqrt{2 D_0}\cdot \begin{cases}
1 & |x|<x_c \\
\alpha \left|\frac{x}{\ell}\right|^{1-1/\alpha} & |x|\geq x_c
\end{cases}
\label{eq:DiffMulti}
\end{equation}
where $x_c=\ell \alpha^{\frac{\alpha}{1-\alpha}}$, so $D(x)$ is continuous. %Here $x=0$ is not a problematic point (i.e. neither zero nor pole). Also 
We define the process in $(-\infty,\infty)$. Furthermore, to fulfill the growth condition we demand $\alpha>0$, see \cite{Gardiner} and App.~\ref{App:Growth}. Here, using the transformation $y(x)\equiv \int_0^{x} {\rm d}x' D(x)^{-1/2}$,  we find 
\begin{equation}
\frac{d\sqrt{D(y)}}{dy}\frac{1}{\sqrt{D(y)}}=\begin{cases}
0 & |y|<y_c \\
\frac{-U_0}{|y|} & |y|>y_c,
\end{cases}
\label{eq:27}
\end{equation}
which is the effective force defined in Eq.~\eqref{eq:y}. The concentration of Brownian particles with the effective force Eq.~\eqref{eq:27}, initially on the origin and $-1<U_0<1$, is
\begin{equation}
P(y,t)\approx \begin{cases}
\frac{1}{\Gamma\left(\frac{1}{2}-\frac{U_0}{2}\right)}|y|^{-U_0}(2t)^{\frac{U_0}{2}-\frac{1}{2}}e^{-\frac{y^2}{2t}} & |y|>y_c \\
\frac{1}{\Gamma\left(\frac{1}{2}-\frac{U_0}{2}\right)}(y_c)^{-U_0}(2t)^{\frac{U_0}{2}-\frac{1}{2}} & |y|<y_c,
\end{cases}
\end{equation}
when the long-time limit is taken, see derivation in \cite{dechant2011solution}.
Back to $P(x,t)$ we obtain
\begin{eqnarray}
P(x,t)&& \approx \frac{1}{\Gamma\left[\frac{1-U_0}{2}\right]}\frac{\ell ^{-U_0+\frac{U_0}{\alpha}}}{(4 D_0 t)^{({1-U_0})/{2}}} \cdot
\label{eq:PxtRegularized}
\\ \nonumber
&&\begin{cases}
\frac{1}{\alpha}\left|\frac{x}{\ell}\right|^{-1+\frac{1}{\alpha}} |x|^{-\frac{U_0}{\alpha}} \exp\left[-\frac{\ell^{2-\frac{2}{\alpha}}x^{\frac{2}{\alpha}}}{4 D_0 t}\right], &  |x|\geq x_c, \\
x_c^{-\frac{U_0}{\alpha}}, & |x|<x_c.
\end{cases}
\end{eqnarray}
The infinite density, given by the condition $J =0$, is
\begin{eqnarray}
&&{\cal I}_{\infty}(x)= C D(x)^{-\frac{A}{2}}=
\label{eq:ID_Regularized}
\\ \nonumber
&& \frac{C}{(2D_0)^{A/2}} 
\begin{cases}
|\alpha|^{-A}\left|\frac{x}{\ell}\right|^{A(-1+\frac{1}{\alpha})}, &  |x|\geq x_c, \\
1, & |x|<x_c
\end{cases}
\end{eqnarray}
so the relation
\begin{equation}
\lim_{t \rightarrow\infty }P(x,t)t^{\frac{\alpha}{2}-\frac{A}{2}(\alpha-1)}  = {\cal I}_{\infty}(x) 
\label{eq:InfinitePDFREgularized},
\end{equation}
holds, similarly to Eq.~\eqref{eq:TimeDependentItoHanggiLimit}, with
$
C=2^{-\alpha(1-A)-\frac{A}{2}}D_0^{-{\alpha(1-A)}/{2}}\ell^{-U_0}|\alpha|^{A-1}/{\Gamma\left[\frac{1-U_0}{2}\right]}
$. %The infinite density coincides the time-dependent solution in the region $x \ll (D_0 t)^{\alpha/2}\ell ^{\alpha-1}$.
As was mentioned above, from the existence of a nonnormalizable solution ${\cal I}_{\infty}(x)$ related to $P(x,t)$, as given in \eqref{eq:InfinitePDFREgularized},  one can prove that Eqs.~\eqref{eq:ZetaGoToOne} and \eqref{eq:zeta} with $\beta$ given in Eq.~\eqref{eq:beta} still hold, so infinite ergodic theory is valid. The proof and the results are similar to Sec.~\ref{sec:PurePowerLawInfinite Ergodic Theory}.
From Eq.~\eqref{eq:ID_Regularized} we see that the nonnormalized state has a structure, which deviates
from a pure power law Eq.~\eqref{eqLSteadyState}. Generally, since ${\cal I}_{\infty}(x) \propto D(x)^{-A/2}$, the infinite density
 is specific to the details of the system.

\subsection{Ergodic Phase}
In Eq.~\eqref{eq:DiffMulti} we have regularized $D(x)$ in the vicinity of the origin [compare with Eq.~\eqref{eq:Diff}], namely $D(x)\neq 0$ when $x\rightarrow 0$. Therefore, when $U_0 > 1$, one finds a normalizable steady state and the process is ergodic, see \cite{dechant2011solution,bray2000random}. The equilibrium distribution is given by 
\begin{eqnarray}
&&P_{\rm eq}(x)= \\
&& \frac{1-U_0}{2x_c(1-U_0-\alpha)} 
\begin{cases}
|\alpha|^{-A}\left|\frac{x}{\ell}\right|^{A(-1+\frac{1}{\alpha})}, &  |x|\geq x_c, \\
1, & |x|<x_c
\end{cases} \nonumber
\end{eqnarray}
which is now normalized as usual, i.e. $\int_{-\infty}^{\infty}P_{\rm eq}(x){\rm d}x =1$, and the standard ergodic theory holds. It means that in the long time limit, when ${\cal O}[x(t)]$ is integrable with respect to the equilibrium state, one finds 
\begin{equation}
{\rm P}(\xi) = \delta(\xi-1),
\end{equation}
where here $\xi \equiv \lim_{t\rightarrow\infty}{\overline{\cal O}}/{\langle {\cal O}\rangle } $. Interestingly, this ergodic phase is obtained when using It\^o interpretation with $\alpha > 2$. For other forms (i.e. Stratonovich or H\"anggi-Klimantovich) with $\alpha >0$ the ergodic phase cannot be obtained with Eq.~\eqref{eq:LangevinMulti}. Needless to say that in this case, when adding a binding force to the Langevin equation \eqref{eq:LangevinMulti}, e.g. an harmonic potential, one may obtain an ergodic phase, with all interpretations.

\subsection{Simulation results}
\subsubsection{Scaled Time-Dependent Solution Approaches the Infinite Density - Stratonovich}
\begin{figure}
\includegraphics[trim = 90 250 90 270, width=\columnwidth]{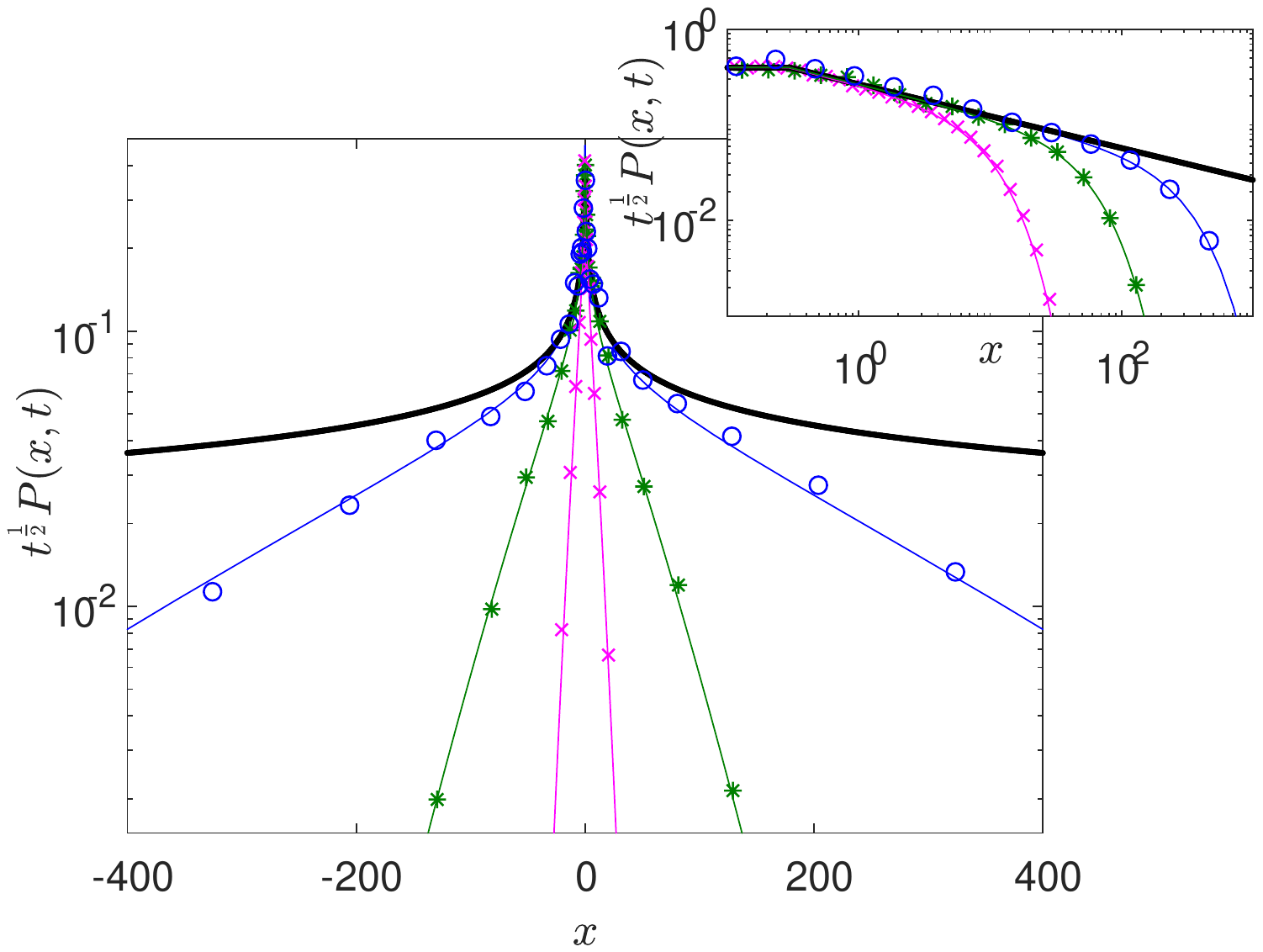}
\caption[$P(x,t)t^{1/2}$ in a multiplicative noise process]{The scaled concentration $P(x,t)t^{1/2}$, where $D(x)$ is given in Eq.~\eqref{eq:DiffMulti} with $\alpha=3/2$, at times; $t=10$ (pink crosses), $t=10^2$ (green stars), and $t=10^3$ (blue circles). The colored lines and the black line represent the analytic expressions $P(x,t)$ [Eq.~\eqref{eq:PxtRegularized}] and ${\cal I}_\infty(x)$ [Eq.~\eqref{eq:ID_Regularized}], respectively. Inset: the data is presented in double-log scale, so the collapses at small $x$ and the deviations from ${\cal I}_{\infty}$ at large $x$ are visible (only positive $x$ is presented).  For ensemble averaging we use $10^4$ realizations.     }
\label{fig:MultiNoiseCollapseSmall}
\end{figure}

%Specifically, %to avoid singularities; poles or zeros, in the diffusion coefficient 
%in this section we choose
Consider the Langevin equation \eqref{eq:LangevinMulti} with the  spatially dependent diffusion coefficient 
% \begin{eqnarray}
% \sqrt{D(x)}=\sqrt{2 D_0}\cdot\begin{cases}
% 1 & |x|<x_c \\
% \alpha \left|\frac{x}{\ell}\right|^{1-1/\alpha} & |x|\geq x_c
% \end{cases}
% \label{eq:DiffMulti}
% \end{eqnarray}
Eq.~\eqref{eq:DiffMulti} with $\alpha=3/2$
%The constant $D_0$ has units of $[\rm {cm}^{2}{\rm sec}^{-1}]$, and $a$ has unit of $[{\rm cm}^{-1+1/\alpha}]$. 
% Generally the exponents $\alpha$ may be positive or negative. Currently we concentrate on the case where $\alpha>0$ (negative $\alpha$ is studied below).
% We choose the transition point as $x_c= \ell \alpha^{\frac{\alpha}{1-\alpha}}$, so the diffusion coefficient is continuous. 
%For simplicity we use in all the simulations in the paper 
and the initial position of all particles is on the origin, i.e. $P(x,t)|_{t=0}=\delta(x)$.
In our simulations of the concentration we use the Langevin equation with Stratonovich interpretation. Here $\beta=1/2$ (see definition in Eq.~\eqref{eq:beta}). 
In Fig.~\ref{fig:MultiNoiseCollapseSmall} we present $P(x,t)t^{1/2}$ versus $x$ for several times. The data collapse on a single curve is found for small $x$ since then it merges with ${\cal I}_\infty(x)$ which is a time-independent state.% In Fig.~\ref{fig:MultiNoiseCollapseLarge} we presents the results of the scaled PDF $P(\chi,t)$ for several time.  

% \begin{eqnarray}
% &&P(y,t)= \label{eq:ScaleMulti}\\ \nonumber
%  && = \begin{cases}
%   \frac{1}{\sqrt{2\pi}} \exp\left[{-y^2/2}\right],  \ \ \ \ \ \ \ \ \ \ \ \ \ \ \ \ \   |y|>\frac{x_c^{1/\alpha}\ell ^{1-1/\alpha}}{\sqrt{2D_0t}},\\
%     \frac{\alpha}{\sqrt{2\pi}}\left(\frac{\ell^2}{2D_0 t y^2 }\right)^{\frac{1-\alpha}{2}}\exp\left[{-\frac{y^{2\alpha}}{2}\left(\frac{\ell^2}{2D_0 t}\right)^{1-\alpha}}\right],  \\  \ \ \ \ \ \ \ \ \ \ \ \ \ \ \ \ \ \ \ \ \ \ \ \ \ \ \ \ \ \ \ \ \ \ \ \ \ \ \ \ \  |y|< \frac{x_c^{1/\alpha}\ell ^{1-1/\alpha}}{\sqrt{2D_0t}},
%   \end{cases}
% \end{eqnarray}

% \begin{equation}
% P(y,t)=
%   \frac{e^{-y^2/2}}{\sqrt{2\pi}}   
%  \qquad {\rm when\ } y=\frac{|x|^{1/\alpha}{\rm sign}(x)}{\sqrt{t}}.
% \label{eq:ScaleMulti}
% \end{equation}

%  Here we find two limiting behaviors. The one is the infinite density Eq.~\eqref{eq:PDF}, and the second Eq.~\eqref{eq:ScaleMultiLim} which is a scaling solution. 
% Thus the process is described by two complementary limit distributions. The basic idea of the concept is that the order of limits is important; whether taking $t\rightarrow\infty$ and fix $x$ and get Eq.~\eqref{eq:PDF}, or fix $x/t^{\alpha/2}$ (equivalent to fixed $y$) and then taking long $t$, (i.e. long $x$ is considered as well) and get Eq.~\eqref{eq:ScaleMultiLim}.
% This double limiting behavior is related to the strong anomalous diffusion as is discussed in the following section. 

\subsubsection{Infinite Ergodic Theory and Ergodic Phase}
\begin{figure}
\includegraphics[trim= 100 280 100 300, width=\columnwidth]{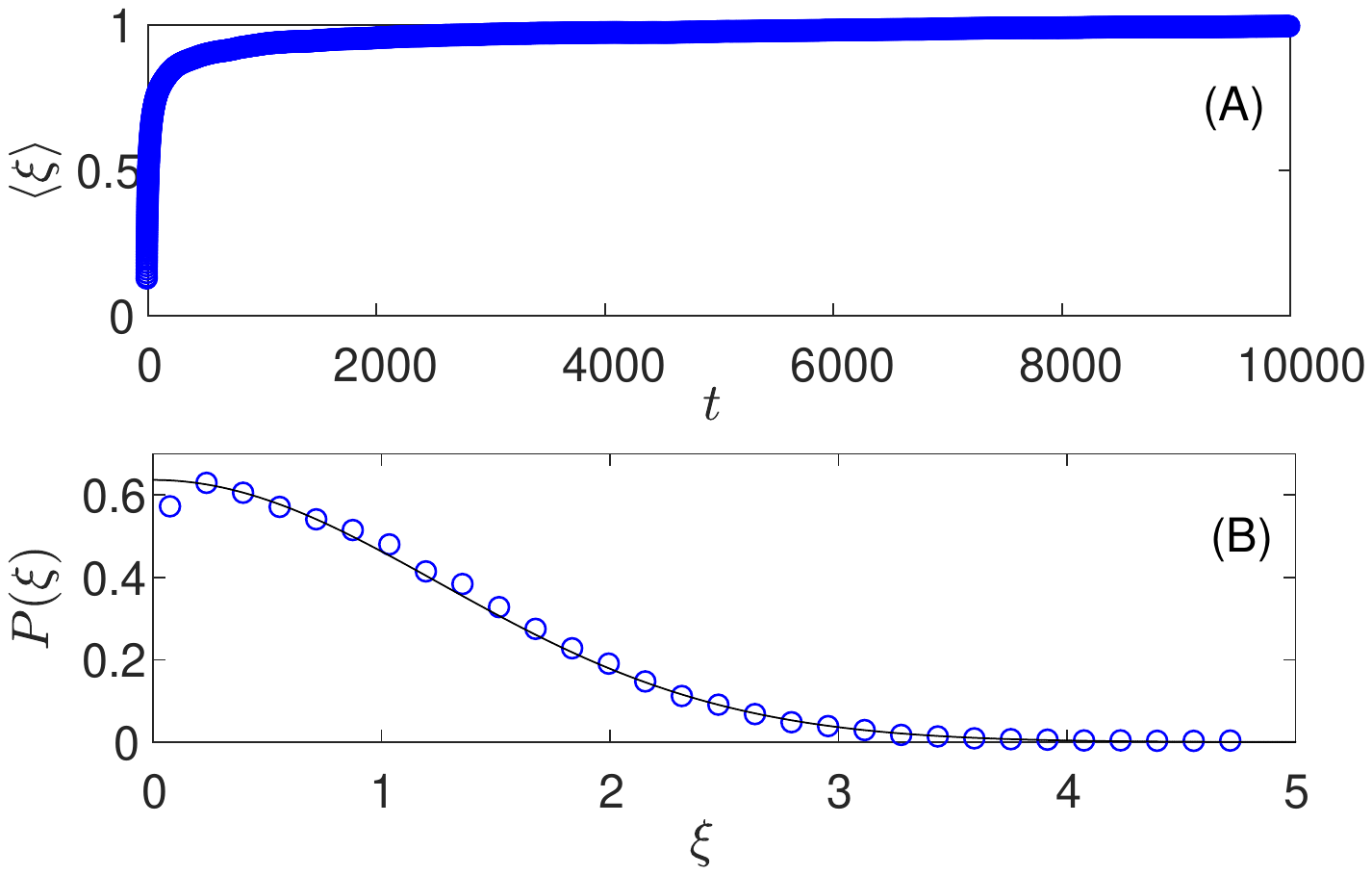}
\caption[$\langle \xi\rangle $ and the distribution of $\xi$]{Simulation of $\xi$ defined in Eq.~\eqref{eq:ZetaDefine} with ${\cal O}[x(t)]=\theta(|x(t)|<10)$ and $D(x)$ is given in Eq.~\eqref{eq:DiffMulti}. Panel (A): The convergence of $\langle  \xi \rangle $ to $1$ in the long time limit. Panel (B):  The distribution of the random variable $\xi$. The data from simulation is presented with open circles and the analytic curve Eq.~\eqref{eq:xi} with a solid line. Here we use $t=10^4$, $\alpha=3/2$, and $5\cdot 10^4$ realizations.    }
\label{fig:xiPositiveAlpha}
\end{figure}

We consider the observable  
$
{\cal O}[x(t)]=\theta(|x(t)|<x^*),
$
which means that ${\cal O}[x(t)]$ is the indicator function. %, which is 1 if the particle is in the interval $(-x^*,x^*)$, and is zero otherwise. % The occupation time $T^+_t$ in the interval $(-x^*,x^*)$ up to time $t$ is
We investigate the random variable
% \begin{equation}
% \zeta = \lim_{t\rightarrow\infty}\frac{T^+_t}{\langle T^+_t \rangle }=\lim_{t\rightarrow\infty}\frac{\overline{{\cal O}}_{t^{1/2}}}{2 \langle {\cal O}\rangle _{P_\infty}}=\lim_{t\rightarrow\infty}\frac{\int_0^t {\rm dt'} {\cal O}[x(t')]}{2 t^{1/2} \langle {\cal O}\rangle _{P_\infty}}. 
% \end{equation}
$\xi$ Eq.~\eqref{eq:ZetaDefine}.
% \begin{equation}
% \xi =\lim_{t\rightarrow\infty}\frac{\frac{1}{t}\int_0^t {\rm dt'} {\cal O}[x(t')]}{2  \langle {\cal O}\rangle }. 
% \label{eq:zetaRegularizedPositive}
% \end{equation}
% The observable ${\cal O}(t)$ define a two-state model. From the renewal model point of view, the sojourn time outside the interval $(-x^*,x^*)$ is fat tailed distributed where $\psi_{\rm off}(\tau)\propto \tau^{-3/2}  $ and the sojourn time in the given interval is exponentially distributed. Thus, from renewal theory, we obtain the distribution of $\xi$ in the long time limit
With the Stratonovich interpretation we expect 
\begin{equation}
{\rm P}(\xi) = {\cal M}_{1/2}(\xi)\equiv \frac{2}{\pi}\exp\left(-\frac{\xi^2}{\pi}\right),
\label{eq:xi}
\end{equation}
i.e. the Mittag-Leffler distribution of order $1/2$ which is one-sided Gaussian where clearly $\langle \xi \rangle =1$.
In the simulation results presented in Fig.~\ref{fig:xiPositiveAlpha} we choose $\alpha=3/2$ and $x^*=10$. We compare between $\langle\overline{{\cal O}}_t\rangle$ from the simulation and the analytic prediction $2 t^{1/2}\langle {\cal O} \rangle \approx 7.17$.  % and we show that $ \langle \overline{{\cal O}}_t\rangle$ converges to $2\langle {\cal O} \rangle \approx 7.17 t^{-1/2}$ in the long time limit. 
To be precise we show that $\langle \xi \rangle \rightarrow 1$ in the long time limit. Furthermore, we show that the distribution of $\xi$ follows the one-sided Gaussian distribution Eq.~\eqref{eq:xi} as expected, also for finite time simulations.

\begin{figure}
\includegraphics[trim= 100 260 110 270, width=\columnwidth]{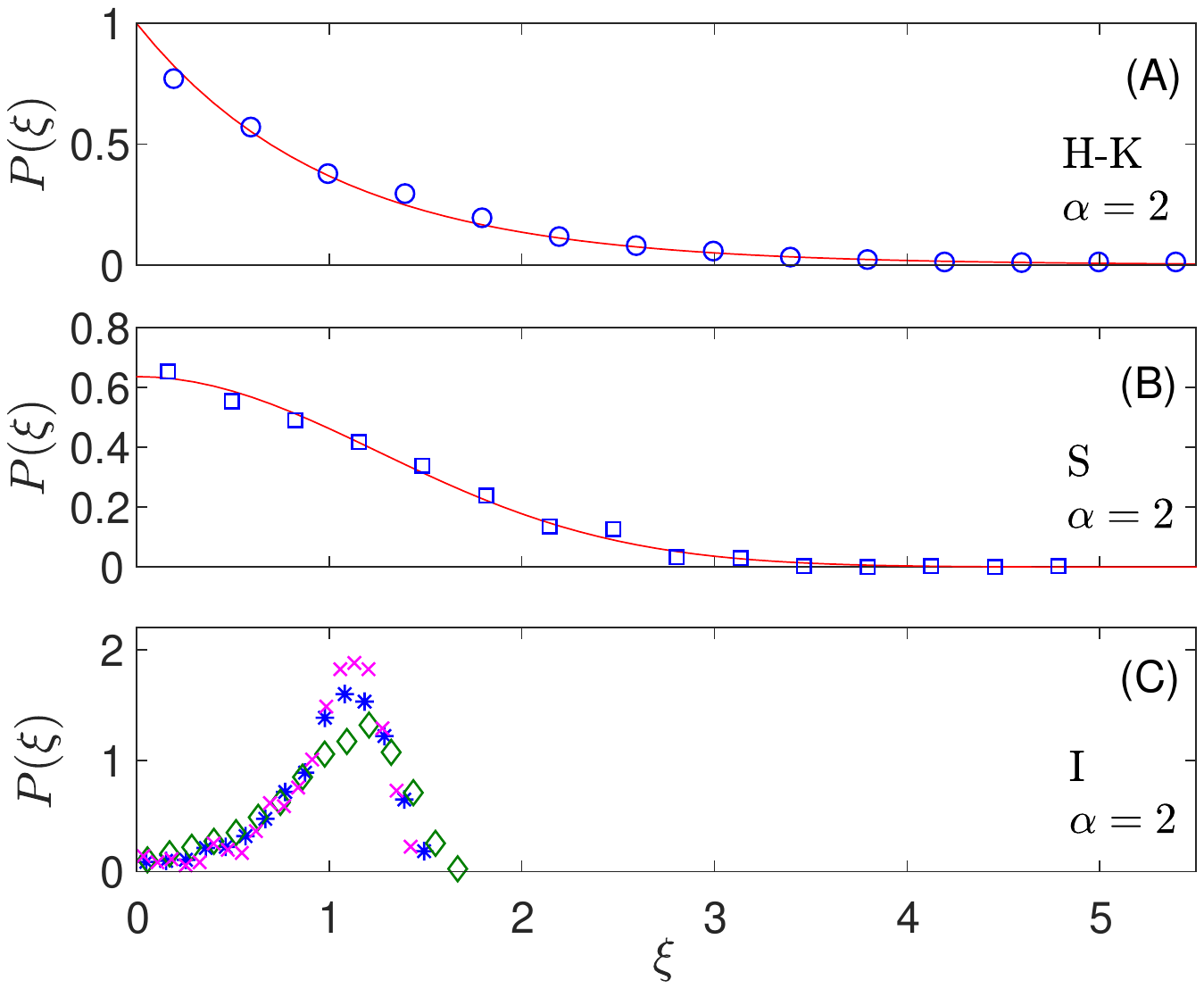}
\caption{The PDF ${\rm P}(\xi)$ for $\xi$ defined in Eq.~\eqref{eq:ZetaDefine} with ${\cal O}[x(t)]=\theta(|x(t)|<10)$.  Here $\sqrt{D(x)}$ is given in Eq.~\eqref{eq:DiffMulti} with $\alpha=2$. 
The data is from $10^4$ realizations. Notice that all panels share the same x-axis. The symbols represent the data from simulations: H\"anggi-Klimontovich (panel (A), $t=10^5$, blue circles), Stratonovich [panel (B), $t=10^5$, blue  rectangles], and It\'o [panel (C), $t=10^3$ (green diamonds), $t=10^5$ (blue stars), $t=10^6$ (pink crosses)]. In panels (A) and (B) we also compare simulations with the analytic predictions given in Eq.~\eqref{eq:Alpha2}  (red solid curves). For the It\'o interpretation, shown in panel (C), ${\rm P}(\xi)$ approaches a delta function in the long time limit. %The deviation form this analytic prediction is a finite time effect, though clearly, while increasing the measurement time the distribution becomes narrower.    
}
\label{fig:P_xi_alpha2}
\end{figure}

In addition, in Fig.~\ref{fig:P_xi_alpha2} we present the simulation results, with $\sqrt{D(x)}$  given by Eq.~\eqref{eq:DiffMulti} and $\alpha=2$. Then, the PDFs of $\xi$, presented with solid curves, are 
\begin{eqnarray}
{\rm P}(\xi)_{\alpha \rightarrow 2}=\begin{cases}
\exp(-\xi), & {\rm HK},
\\
\frac{2}{\pi}\exp\left(-\frac{\xi^2}{\pi}\right), & {\rm S},
\\
\delta(\xi-1), & {\rm I}.
\end{cases}
\label{eq:Alpha2}
\end{eqnarray}
The derivation of these results is given in App.~\ref{AppD}. We compare these analytical results with the simulations when $t=10^5$, and $10^4$ particles.   For the It\'o interpretation the deviation from this analytic prediction, presented in Fig.~\ref{fig:P_xi_alpha2} panel (C), is a finite time effect, since while increasing the measurement time the distribution becomes narrower. We see that the statistics of time averages clearly depend on the interpretation.  

% \section{It\'o and H\"anggi-Klimantovich Interpretations to Eq.~(1). }
% As was mention, Langevin equation with a multiplicative noise, may take different interpretations, in relation to the process it is described. 

% We start with studying Eq.~\eqref{eq:LangevinMulti} where $x(t)\in[0,\infty)$ and $\alpha$ is positive. 

\section{Bounded Processes}
\label{sec:negative}
In the previous sections we have studied processes in an infinite domain. 
In many examples, ergodicity is discussed in the context of a finite sized system, simply because thermodynamics 
is valid for systems of finite (though large) size.
Hence, we wish to explore infinite ergodic theory for inhomogeneous diffusion in a finite domain.
In previous section, the infinite size system limited us to the condition of $\alpha > 0$. Here, we examine bounded processes, namely when $x \in [0,L]$.
This allows us to choose negative $\alpha$, provided the growth condition holds, see \cite{Gardiner} and App.~\ref{App:Growth}. As will be shown in the following, the non-integrable point of the zero-current solution ${\cal I}_{\infty}(x)$ is at $x\rightarrow 0$ (instead of $x\rightarrow \infty$ as in the positive $\alpha$ cases above).

 \subsection{Pure Power-law dependent $\sqrt{D(x)}$ with Stratonovich Interpretation} 

For simplicity here we consider only the Stratonovich interpretation. 
The process is bounded in $[0,L]$ and
\begin{equation}
\sqrt{D(x)}=
\sqrt{2D_0}|\alpha| (x/\ell) 	^{1-1/\alpha} {\rm \ when\  } 0 \leq  x \leq L
\label{eq:39}
\end{equation}
with $\alpha<0$. Clearly as $x\rightarrow 0 $ the diffusivity becomes small, and hence intuitively a particle in the vicinity of zero is slowed down.  This, in turn, implies a pile up of particles close to zero, which is associated with the non-integrable state.  
%  AS before, the Fokker-Planck equation may takes different forms refer to the interpretation used.
 % form The Fokker-Planck equation using Stratonovich approach is 
%  \begin{equation} 
%  \frac{\partial P(x,t)}{\partial t}=2D_0   \alpha\frac{\partial }{\partial x}\left\{|x|^{1-\frac{1}{\alpha}}\frac{\partial}{\partial x}\left[|x|^{1-\frac{1}{\alpha}}P(x,t)\right]\right\}.
%  \label{eq:FP}
%  \end{equation}
 Then the infinite density which, as mentioned, is defined via Eq.~\eqref{eq:CurrentDef}, is 
 \begin{equation}
  {\cal I}_{\infty}(x)= C D(x)^{-1/2} =C \frac{\ell ^{1-1/\alpha }}{\sqrt{2D_0\alpha^2}}
  x^{-1+1/\alpha},
  \label{eq:InfiniteDensityNegative}
 \end{equation}
where $C$ is determined below using the time-dependent solution. Here ${\cal I}_{\infty}(x)$ is nonnormalizable due to its behavior close to zero.
 
 To solve the Fokker-Planck equation \eqref{eq:FP_Stratonovich},  we define a new variable
  \begin{equation}
    y(x)\equiv \int_x^{L}\frac{{\rm d}x'}{\sqrt{D(x')}} = \frac{\ell ^{1-1/\alpha}}{\sqrt{2D_0 }}\left[x^{1/\alpha}-L^{1/\alpha}\right],
    \label{eq:42}
  \end{equation}
%     \begin{equation}
%     y(x)\equiv \frac{\ell ^{1-1/\alpha}}{\sqrt{2D_0 }}x^{1/\alpha},
%     \label{eq:42}
%   \end{equation}
%thus the equation for $y$ does not have any additional effective force and the Fokker-Planck equation is 
then Eq.~\eqref{eq:FP_Stratonovich}, with $A=1$ (i.e. Stratonovich form), transforms into
\begin{equation}
\frac{\partial}{\partial t}P(y,t)=\frac{1}{2}\frac{\partial^2}{\partial y^2}P(y,t)
\label{eq:43}
\end{equation}
in the interval 
$ y\in [0,\infty)$  
%$ y\in [L^{1/\alpha},\infty)$ 
with the reflecting boundary condition 
%$\partial_y P(y,t)|_{y=L^{1/\alpha}}=0$.
$\partial_y P(y,t)|_{y=0}=0$. 
%Thus, the equation for $y$ does not have any additional effective force. 
Solving Eq.~\eqref{eq:43}
with the method of images, considering the initial condition $y(t)|_{t=0}=y_0$, and transforming back to $x$ using Eq.~\eqref{eq:42}, gives
\begin{eqnarray}
 P(x,t)=  \frac{(x/\ell)^{-1+\frac{1}{\alpha}}}{\sqrt{4 D_0 \pi t}|\alpha| } 
  \nonumber \left\{
\exp\left[-\frac{\left(x^{\frac{1}{\alpha}}-x_0^{\frac{1}{\alpha}}\right)^2\ell^{2-\frac{2}{\alpha}}}{4 D_0t} \right] \right. \\ 
  \left.+\exp \left[-{\frac{\left(x^{\frac{1}{\alpha}}+x_0^{\frac{1}{\alpha}}-2L^{\frac{1}{\alpha}}\right)^2\ell^{2-\frac{2}{\alpha}}}{4D_0 t }}\right]
\right\}.
\label{eq:PDFNegative}
\end{eqnarray}
The following relation between the time-dependent solution $P(x,t)$ and the infinite density in Eq.~\eqref{eq:InfiniteDensityNegative} is fulfilled
\begin{equation}
\lim_{t\rightarrow\infty} t^{1/2}P(x,t)=\sqrt{\frac{1}{\pi D_0 \alpha^2 }}(x/\ell)^{-1+1/\alpha} = {\cal I}_{\infty}(x),
\label{eq:NegativeSteady}
\end{equation}
hence we identify the constant to be $C=(2\pi)^{-1/2}$. 

\begin{figure}
\includegraphics[trim=  90 280 100 290, width=\columnwidth]{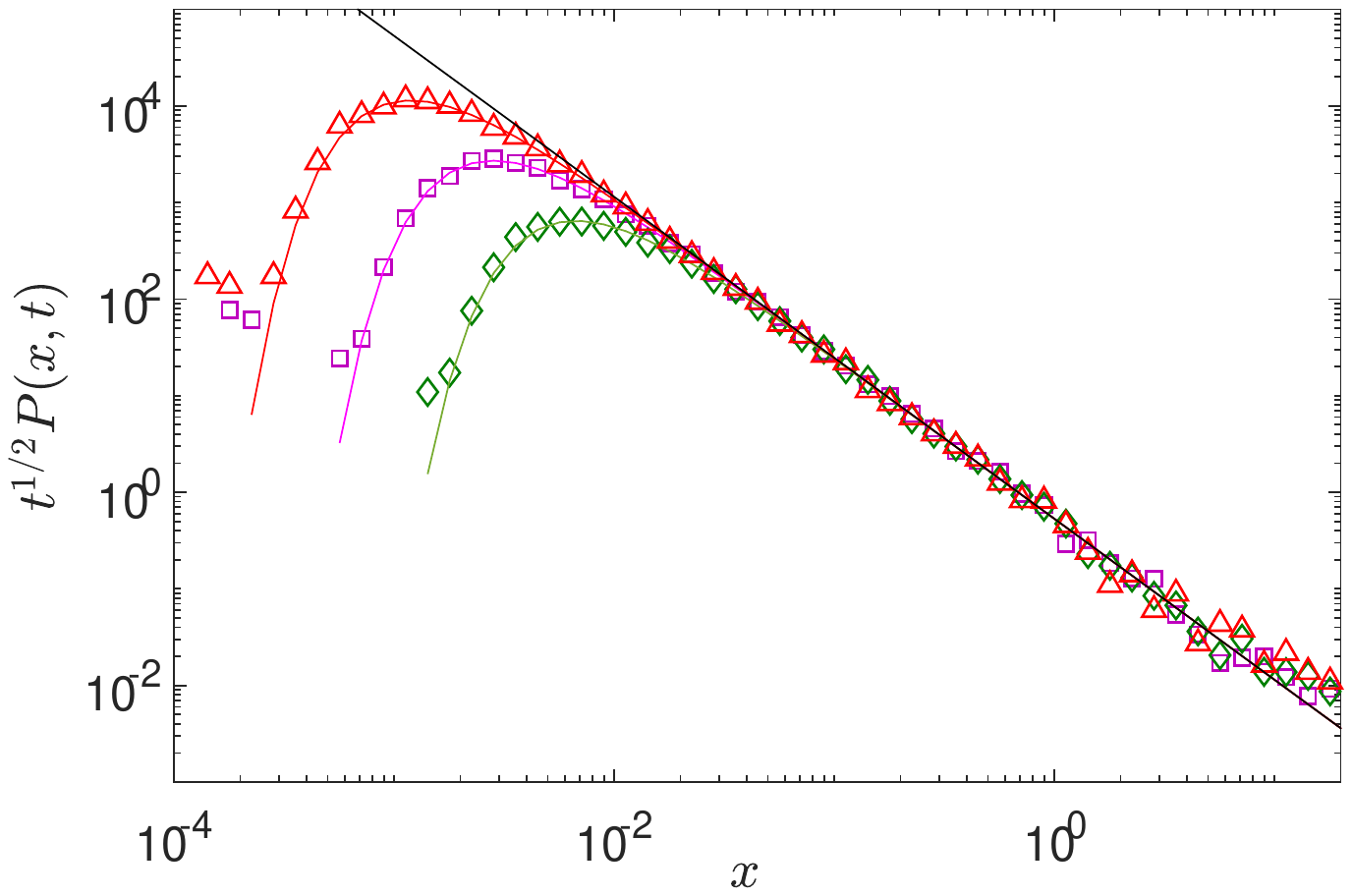}
\caption{The scaled  PDF  $t^{1/2}P(x,t)$ for the process with diffusivity Eq.~\eqref{eq:39}, $\alpha=-1.5$, $L=20$,   and times %$t=100$ (blue circles), 
$t=316$ (green diamonds), $t=1000$ (pink squares) and $t=3162$ (red triangle). The coloured solid curves are Eq.~\eqref{eq:PDFNegative}. The collapse for large $x$ of data to the infinite density [black line, Eq.~\eqref{eq:NegativeSteady}] is clearly visible. The number of particles is $10^4$, and the Stratonovich interpretation is considered. }
\label{fig:PDFNegative}
\end{figure}

In Fig.~\ref{fig:PDFNegative} we present $t^{1/2}P(x,t)$ versus $x$ for several times. The symbols are the simulation results and the solid curves represent the analytic prediction. The collapse of the data for large $x$ and the approach to ${\cal I}_{\infty}(x)$ while increasing the time are clearly visible.

\subsubsection{Infinite Ergodic Theorem}
\label{Sec:text1}
\begin{figure}
\includegraphics[trim= 100 200 100 290, width=\columnwidth]{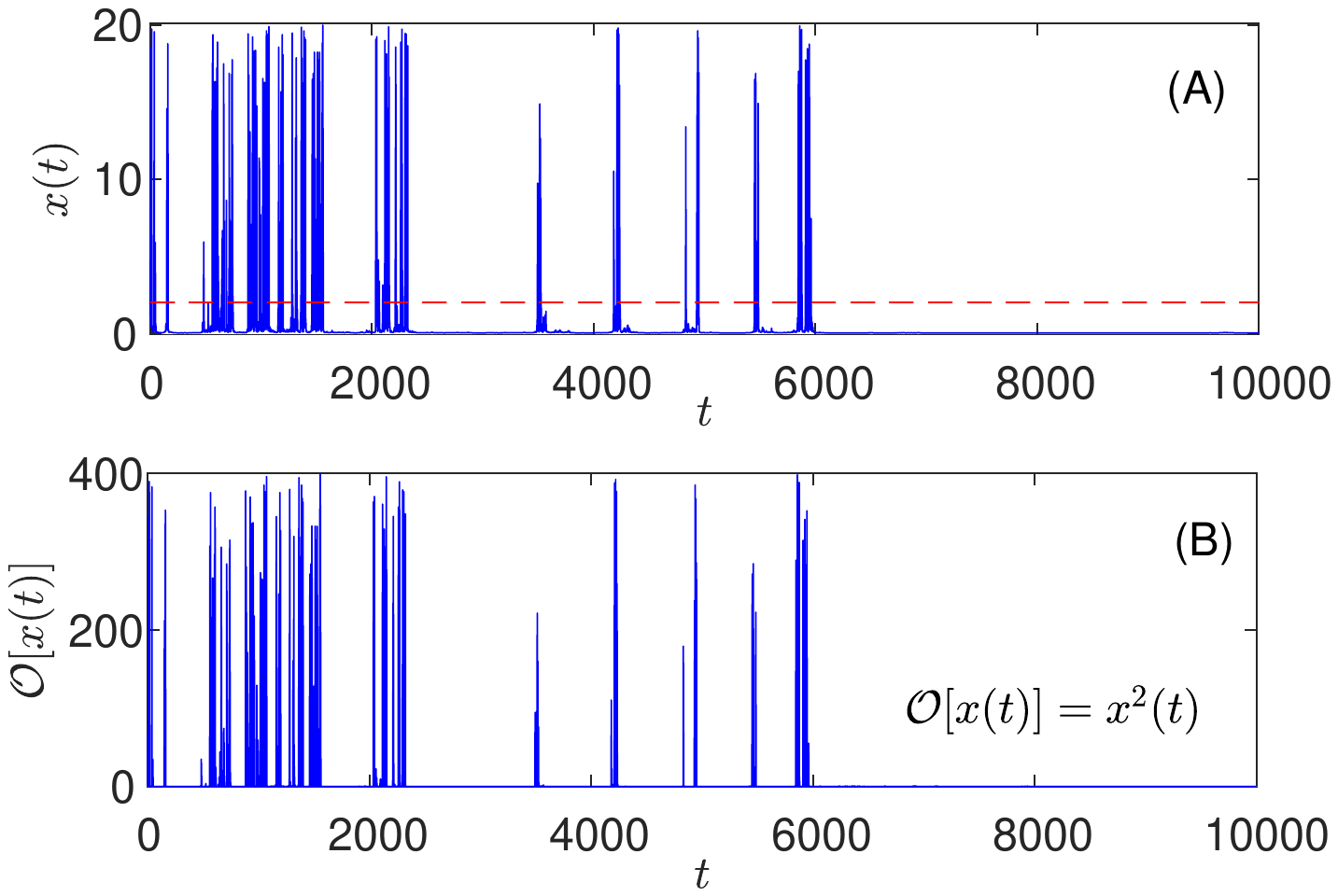}
\includegraphics[trim= 130 70 130 140, width=\columnwidth]{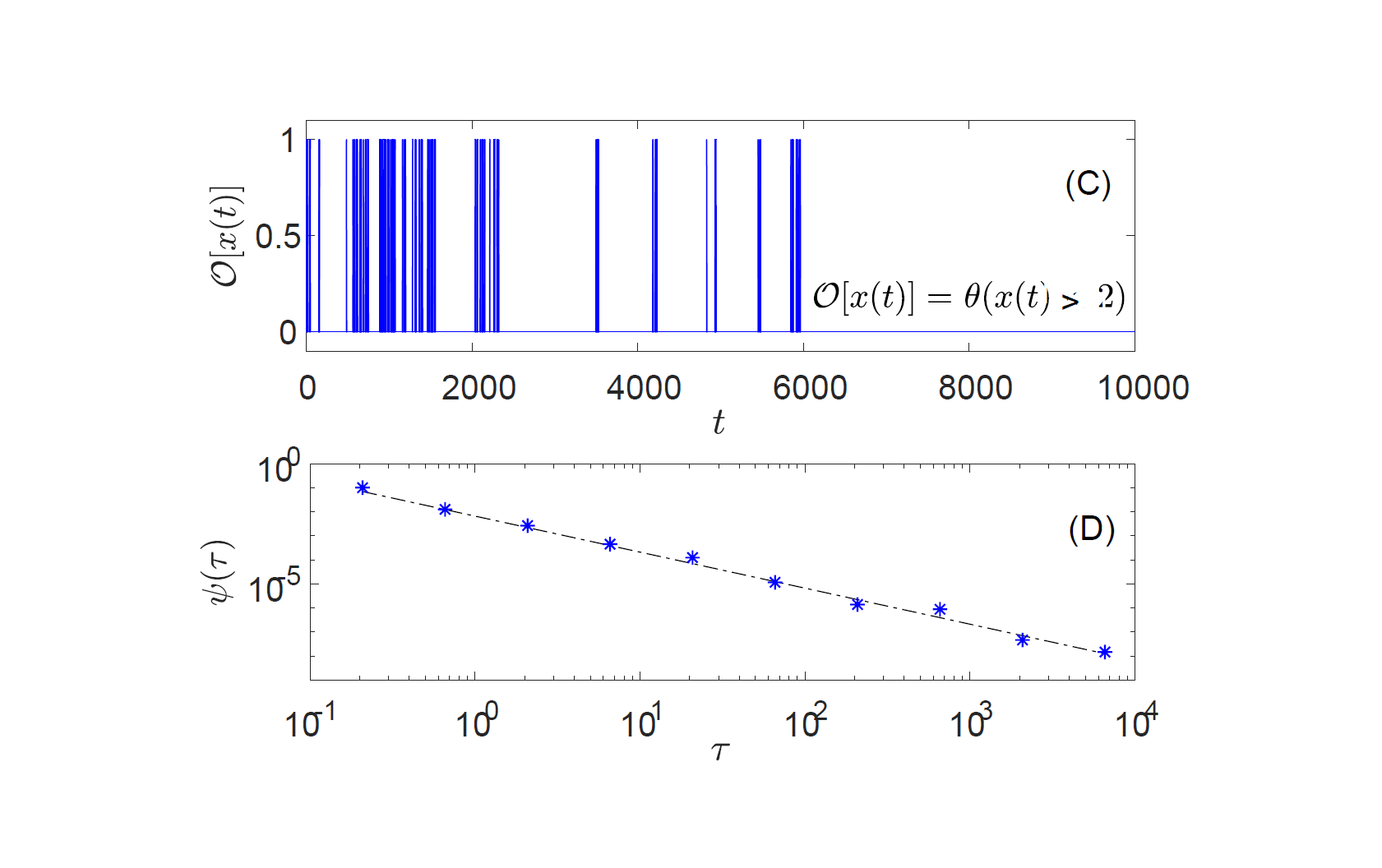}
\caption{Panel (A): A single realization $x(t)$ where $\sqrt{D(x)}$ is given in Eq.~\eqref{eq:39} with  $\alpha=-3/2$ and $L=20$. The process is now bounded in $[0,20]$, still infinite ergodic theory holds. A threshold on $x=2$ is represented with a red dashed line.  Panels (B) and (C): The observable ${\cal O}[x(t)]=x^2(t)$ and ${\cal O}[x(t)]=\theta(x(t)>2)$ respectively. Here, $\theta(x(t)>2)=1$ if $x(t)>2$ and zero otherwise.  The long sojourn times of ${\cal O}[x(t)]$ close to zero is clearly visible.  These are related to the slowdown of diffusion close to $x\rightarrow 0$. In Panel (D) we show the PDF $\psi(\tau)$ of the sojourn times $\tau$ of the trajectory $x(t)$ below the threshold $x=2$. The simulation results are given with blue stars. The black dashed line decreases as $\propto \tau^{-3/2}$, so $\beta=1/2$ in agreement with the analytic prediction,  see \ref{Sec:text1} and Eq.~\eqref{eq:Psi}.    }
\label{fig:trajectoryNegative}
\end{figure}

Consider the observable ${\cal O}[x(t)]=x^2(t)$. This observable is integrable with respect to the infinite density Eq.~\eqref{eq:NegativeSteady} when $\alpha<-1/2$. Thus $x^2(t)$, when plotted versus time, exhibits long sojourn times close to zero, see the illustration in Fig.~\ref{fig:trajectoryNegative}.  The ensemble average of $x^2$ is 
\begin{eqnarray}
\langle x^2\rangle &\stackrel{t \rightarrow \infty}{\approx}& t^{-1/2} \int_0^{L} \sqrt{\frac{1}{\pi D_0 \alpha ^2 }}\frac{x^{1+1/\alpha}}{\ell^{-1+1/\alpha}} {\rm d}x \\ \nonumber
&=& t^{-1/2}\sqrt{\frac{1}{\pi D_0 }}  \frac{L^{2+1/\alpha}}{|2\alpha+1|\ell^{-1+1/\alpha}} .
\end{eqnarray}
In other words the ensemble average is computed with respect to the nonnormalized state ${\cal I}_\infty(x)$, which in that sense replaces the more typical invariant density of the system (when it exists).

\begin{figure}
\includegraphics[trim= 100 280 100 300, width=\columnwidth]{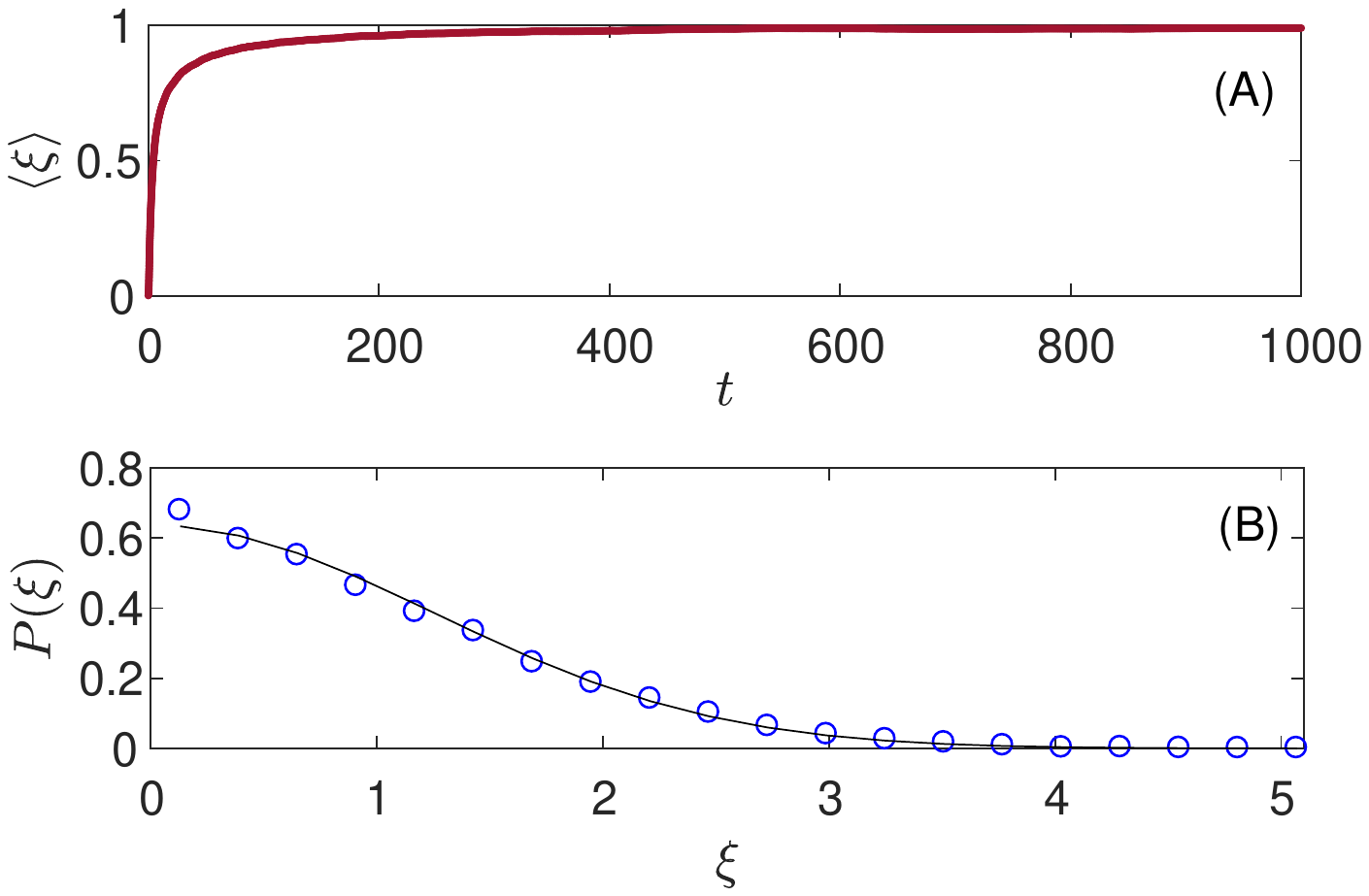}
\caption{Simulating the random variable $\xi$ defined in Eq.~\eqref{eq:ZetaDefine} with the observable ${\cal O}[x(t)]=x^2$ yield the average of $\xi$ and its PDF. The diffusivity is given in Eq.~\eqref{eq:39}, the measurement time is $t=10^3$, $\alpha=-3/2$, $L=20$ and the number of particles is $10^4$. Panel (A) shows the convergence of $\langle \xi \rangle $ to 1 when the time  is long. Panel (B) presents the PDF of $\xi$. The simulation results are given with blue circles, and the analytic prediction Eq.~\eqref{eq:MLNegative} is the black line.     }
\label{fig:TAMSD}
\end{figure}

For the simulations we use $L=20$ and $\alpha=-3/2$, hence $t^{1/2}\langle x^2 \rangle \approx 21.658 $. The random variable $\xi$ defined in Eq.~\eqref{eq:ZetaDefine} with  ${\cal O}[x(t)]=x^2(t)$, using Eq.~\eqref{eq:zeta} with $\beta=1/2$,
% \begin{equation}
% \xi = \lim_{t \rightarrow \infty} \frac{\frac{1}{t}\int_0^t{\rm d}t' x^2(t')}{2\langle x^2 \rangle }
% \end{equation}
is distributed according to
\begin{equation}
{\rm P}(\xi)= {\cal M}_{1/2}(\xi)\equiv\frac{2}{\pi}\exp\left(-\frac{\xi^2}{\pi}\right),
\label{eq:MLNegative}
\end{equation}
i.e. the Mittag-Leffler function of order $1/2$, see Fig.~\ref{fig:TAMSD}. 
%This is known for Brownian particles, which are defined in the process $y$. The mapping $y \rightarrow x$ effectively `compresses' the space in such a way that the temporal properties such as the return times behave similarly.  

\subsection{Normal Diffusion Close to $x=L$ with
Stratonovich Interpretation }
\begin{figure}
\includegraphics[trim= 100 270 100 300, width=\columnwidth]{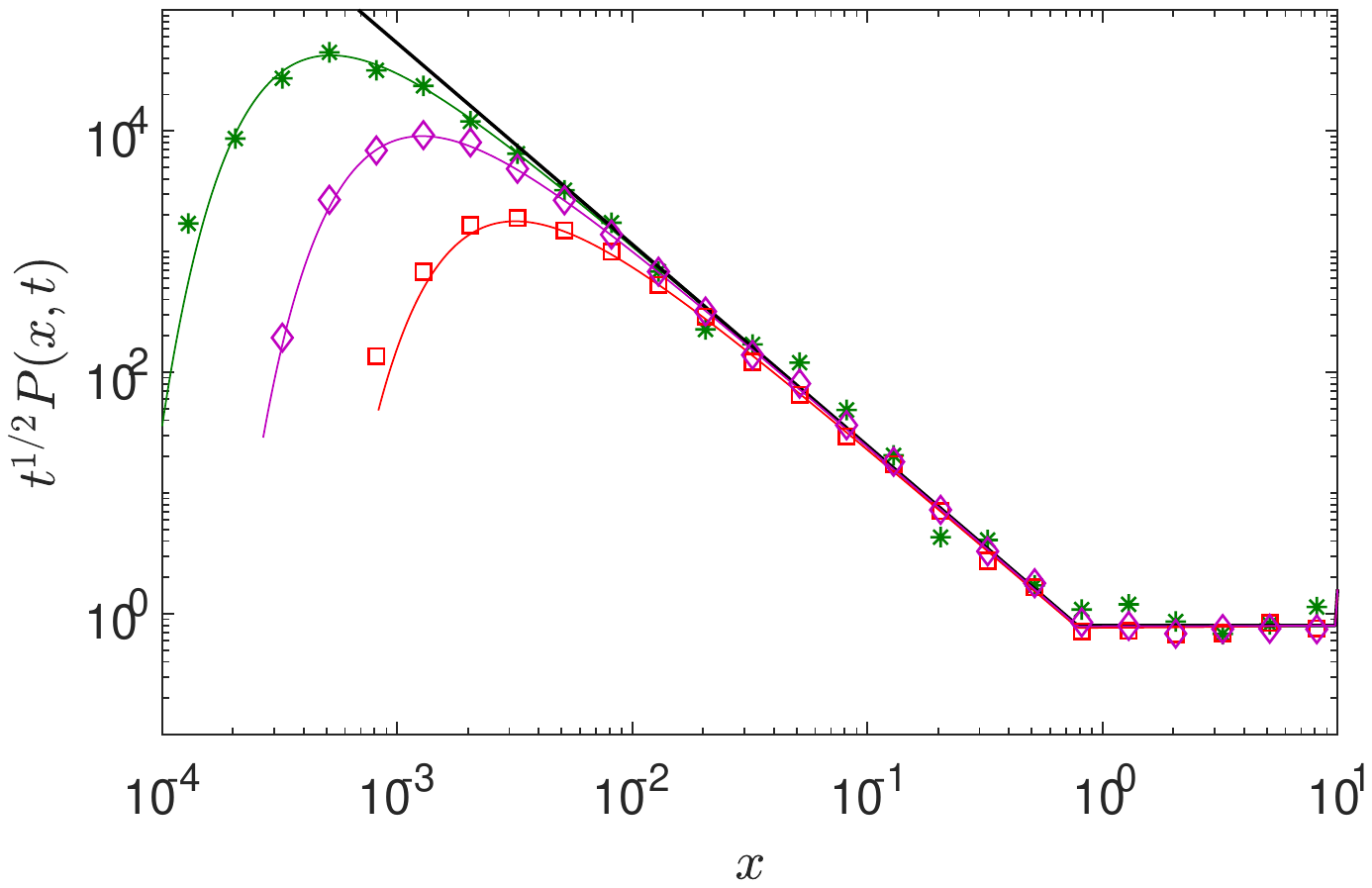}
\caption{The scaled PDF $t^{1/2}P(x,t)$ for different times where $\sqrt{D(x)}=|\alpha|x^{1-1/\alpha}\theta(0\leq x\leq x_c)+\theta(x_c<x\leq L)$ with $P(x,t)|_{t=0}=\delta(x-1/2)$. We use $\alpha=-3/2$ and $L=10$, hence $x_c \approx 0.784$.  We present the simulation results for % $t=316$ (blue circles), 
$t=10^3$ (red triangles), $t=3162$ (pink diamonds) and $t=10^4$ (green stars). The lines represent the analytic prediction Eq.~\eqref{eq:PDFExample1} (with respect to the different times) and the black solid line is Eq.~\eqref{eq:ID(x)Exmqple1}.  The number of particles is $5000$ (except for $t=10^4$ where $10^3$ realizations were taken). Here the infinite density has a clear structure, beyond the simple power law presented in Fig.~\ref{fig:PDFNegative}.
     }
\label{fig:P_x_t_NEW}
\end{figure}

In the previous example ${\cal I}_\infty(x)$ decays as a power law in the whole domain $[0,L]$ peaking on $x=0$. We now consider  
\begin{eqnarray}
&&\sqrt{D(x)}=\\ \nonumber
&&\sqrt{2D_0}\left[|\alpha|\left(\frac{x}{\ell}\right)^{1-1/\alpha} \theta (0\leq x\leq x_c)+\theta (x_c<x\leq L)\right]
\label{eq:RegularDiffusivityNegative}
\end{eqnarray} 
where $\alpha$ is negative. $x_c$ is chosen so $D(x)$ is continuous. Similar to the previous example, $D(x)$ vanishes when $x\rightarrow 0$, but the field $D(x)$ has some structure.
Also here we define  $
    y(x)=\int_x^{L}{{\rm d}x'}\left[{D(x')}\right]^{-1/2}
  $ and find the time dependent solution 
\begin{eqnarray}
 \nonumber P(x,t)= && \frac{1}{\sqrt{2 \pi t D(x)} }  \left\{
\exp\left[-\frac{\left(y(x)-y(x_0)\right)^2}{2 t} \right] \right. \\ 
&&  \left.  +\exp \left[-{\frac{\left(y(x)+y(x_0)\right)^2}{2 t }}\right]
\right\}.
\label{eq:PDFExample1}
\end{eqnarray}
The solution obtained from $J =0$, see Eq.~\eqref{eq:CurrentDef}, is
\begin{equation}
{\cal I}_{\infty}(x)= \sqrt{\frac{1}{\pi D_0 }}
\begin{cases}
\frac{1}{|\alpha|}\left(\frac{x}{\ell}\right)^{-1+1/\alpha} & 0<x<x_c \\
1 & x_c<x<L
\end{cases}
\label{eq:ID(x)Exmqple1}
\end{equation}
which diverges due to its behavior in $x=0$. Here, $\lim_{t\rightarrow\infty} t^{1/2}P(x,t)={\cal I}_{\infty}(x)$. In Fig.~\ref{fig:P_x_t_NEW} we show $t^{1/2}P(x,t)$ versus $x$ which approaches ${\cal I}_{\infty}(x)$ as we increase $t$. Clearly, it illustrates that the structure of the infinite density ${\cal I}_{\infty}(x)$ depends on $D(x)$.

Now we examine an observable  which is integrable in respect to ${\cal I}_{\infty}(x)$, e.g. the mean-square-displacement (MSD) ${\cal O}[x(t)]=x^2$ with $\alpha<-1$. Then the time-averaged MSD is controlled by the details of the nonnormalizable density ${\cal I}_{\infty}(x)$, since
\begin{eqnarray}
&\langle \overline{ x^2(t)} \rangle &= \frac{1}{t}\int_0 ^{t} {\rm dt'}\langle x^2(t') \rangle = \frac{1}{t} \int_0^{t}{\rm d}t' \int_0^L {\rm d}x P(x,t') x^2 \nonumber\\
&\stackrel{t\rightarrow \infty}{\approx}& \frac{1}{t} \int_0^{t}{\rm d}t' \int_0^L{\rm d}x (t')^{-\frac{1}{2}}{\cal I}_{\infty}(x) x^2 \nonumber \\
&=& \frac{t^{-\frac{1}{2}}}{1/2} \sqrt{\frac{1}{\pi D_0 }} \left[ \frac{\ell^{1-\frac{1}{\alpha}}}{(2+1/\alpha)|\alpha|}x_c^{2+\frac{1}{\alpha}}+\frac{L^3}{3}-\frac{x_c^3}{3}\right]  \nonumber \\
&=& 2 \langle x^2 \rangle. 
\end{eqnarray}
From here, if $\alpha=-3/2$ so $x_c \approx 0.784$, and $L=10$ we get $ t^{1/2}\langle x^2\rangle \approx 266.12 $, which is used in Fig.~\ref{fig:mean_xi_x_1-x_} [panel (A)].  We define a variable $\xi\equiv \overline{x^2(t)}/[2\langle x^2(t)\rangle ]$, similar to Eq.~\eqref{eq:ZetaDefine} with $\beta=1/2$, 
% \begin{eqnarray}
% \xi=\frac{\overline{x^2(t')}}{2\langle x^2\rangle },
% \label{eq:zeta1}
% \end{eqnarray}
so $\lim_{t \rightarrow\infty} \langle \xi\rangle =1$, and its PDF $P(\xi)$ follows Eq.~\eqref{eq:xi}.
Fig.~\ref{fig:mean_xi_x_1-x_} presents (with red stars) the simulation results for $\langle \xi \rangle$ [panel (A)] and ${\rm P}(\xi)$ [panel (C)]  with  $\alpha=-3/2$. The measurement time is $t=10^3$ and the ensemble size is $10^4$ particles. The agreement with infinite ergodic theory is visible.

\section{Power-Law Behavior close to more than one point}
In the previous sections we have shown that when the diffusion coefficient has the form Eq.~\eqref{eq:DiffIntroduction} we find a nonnormalizable steady state and infinite ergodic theory holds. These results, for negative $\alpha$, are related to the fact that the spatially dependent diffusion coefficient slows down the particles close to zero. Therefore, similar results are obtained when one chooses diffusion coefficients with more than one ``pathological points'' as is demonstrated in the following. 

\subsection{Example: Two divergent points} %$\sqrt{D(x)}=x(1-x)$  }
\begin{figure}
\includegraphics[trim= 90 270 110 290, width=\columnwidth]{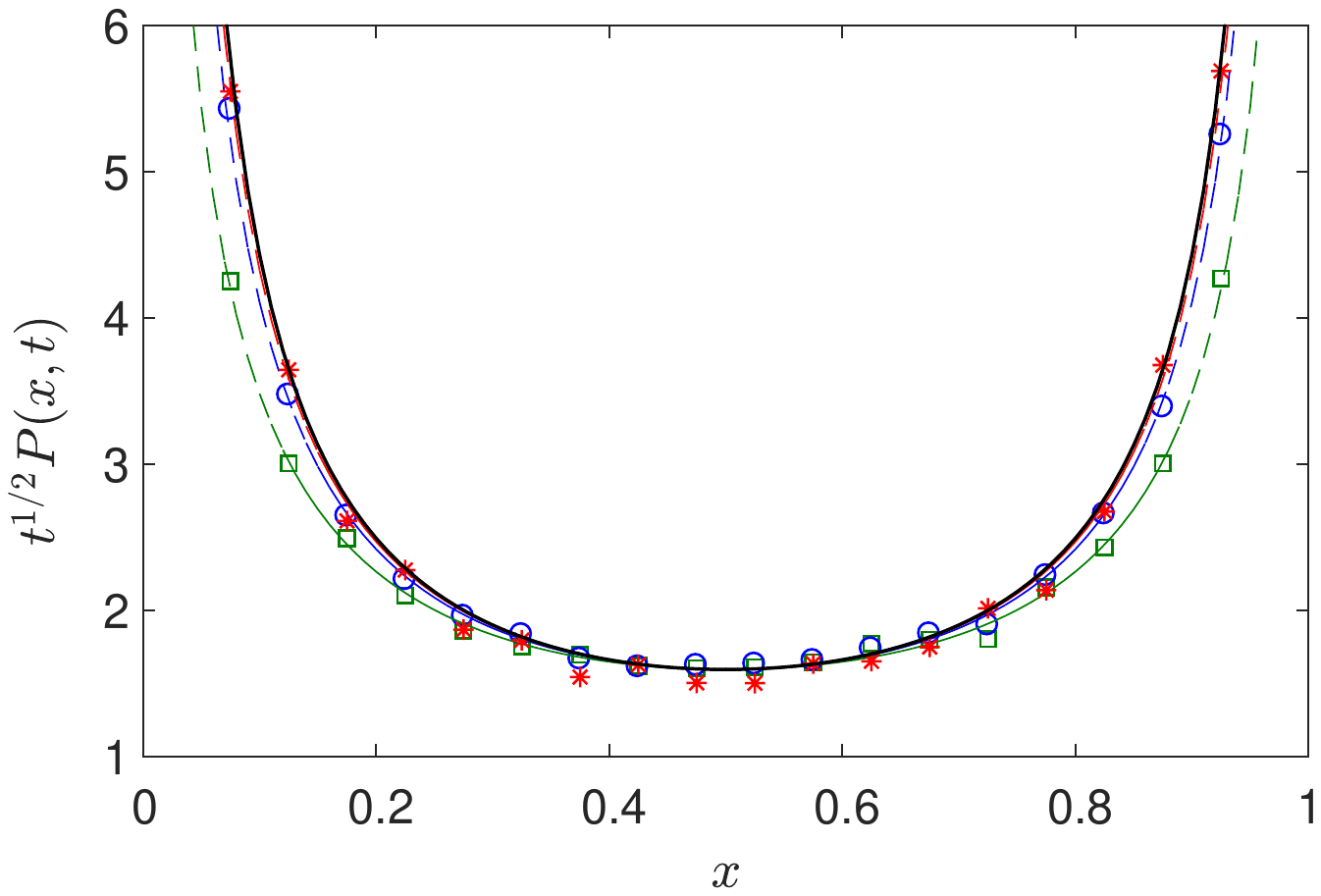}
\caption{ The scaled PDF $t^{1/2}P(x,t)$ for different times where $\sqrt{D(x)}=x(1-x)$ with $P(x,t)|_{t=0}=\delta(x-1/2)$. Here we present the simulation results for %$t=1$ (blue stars), $t=3$ (green crosses),
$t=10$ (green rectangles), $t=31$ (blue dots) and $t=100$ (red stars). The dashed lines represent the analytic prediction Eq.~\eqref{eq:PDF(t)x(1-x)} (with respect to the different times) and the black solid line is Eq.~\eqref{eq:ID(x)x(1-x)}.  The number of particles is $10^5$.}
\label{fig:PDF_x_1-x__new}
\end{figure}

Consider a signal $x(t)$ which evolves via the Langevin equation
\begin{equation}
\frac{dx}{dt}=\sqrt{2D_0}\cdot \frac{x}{L}\left(1-\frac{x}{L}\right)\eta(t)
\label{eq:TwoPointsLangevin}
\end{equation}
where the signal is bounded, i.e.  $0\leq x(t)\leq L$. We use Stratonovich approach and obtain that the time-dependent solution of the Fokker-Planck equation, with initial condition $P(x,t)|_{t=0}=\delta(x-\frac{L}{2})$ and reflecting boundary condition
\begin{equation}
P(x,t)=\frac{L^2}{\sqrt{4 D_0 \pi t} x(L-x)}\exp\left[-\frac{\ln^2\left(\frac{x}{L-x}\right)L^2}{4D_0 t}\right],
\label{eq:PDF(t)x(1-x)}
\end{equation}
which has the form of the log-normal distribution of the variable $x/(L-x)$. The zero-current solution ${\cal I}_{\infty}(x)$, which is obtain from Eq.~\eqref{eq:CurrentDef}, fulfills
\begin{equation}
{\cal I}_{\infty}(x)=\lim_{t\rightarrow\infty} t^{1/2} P(x,t)= \frac{L^2}{\sqrt{4 D_0 \pi} x(L-x)},
\label{eq:ID(x)x(1-x)}
\end{equation}
which diverges due to the boundary points, i.e. $x=0$ and $x=L$.  This result is demonstrated with simulations in Fig.~\ref{fig:PDF_x_1-x__new}. 
Notice that ${\cal I}_{\infty}(x)$ is non-integrable at both $x\rightarrow 0$ and $x\rightarrow L$.

\begin{figure}
\includegraphics[trim= 90 270 90 300, width=\columnwidth]{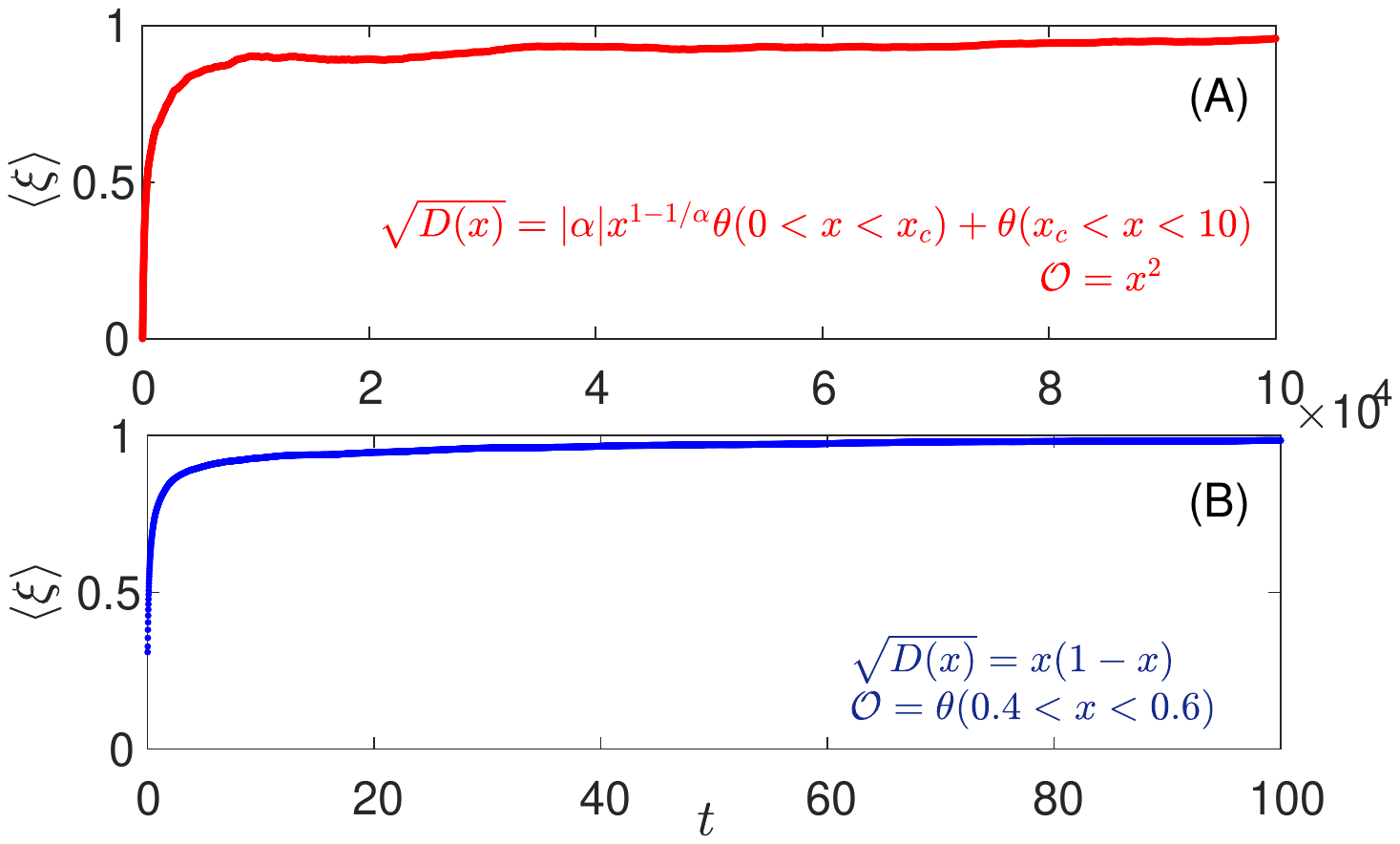}
\includegraphics[trim= 80 270 90 300, width=\columnwidth]{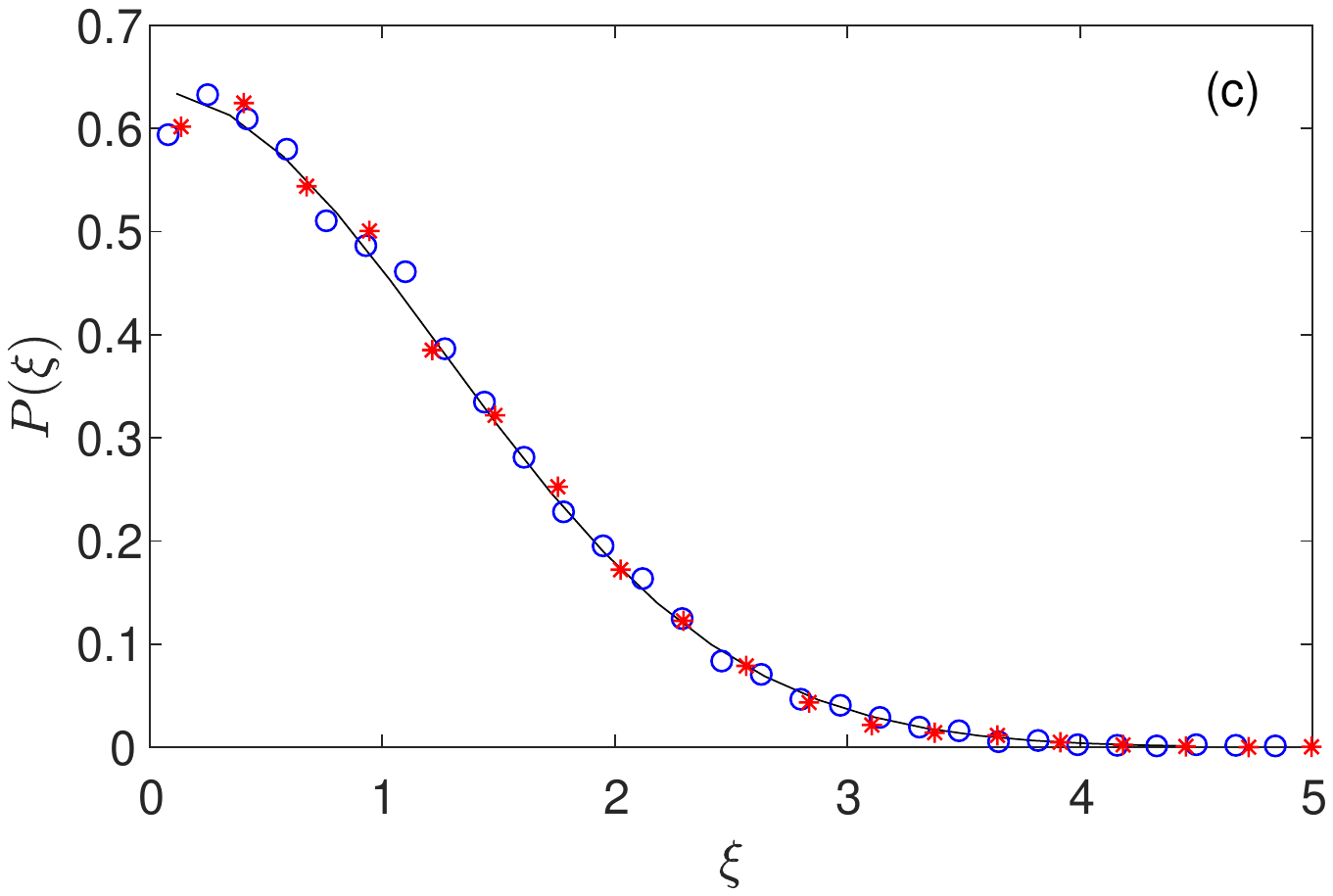}
\caption{Panels (A) and (B): The convergence of $\langle  \xi \rangle  $ to $1$, where $\xi$ are defined in Eq.~\eqref{eq:ZetaDefine} with the observable ${\cal O}[x(t)]=x^2(t)$ and the diffusivity given in Eq.~\eqref{eq:RegularDiffusivityNegative}  (panel (A), red) and ${\cal O}[x(t)]=\theta(0.4<x(t)<0.6)$ for the process define in Eq.~\eqref{eq:TwoPointsLangevin}. (panel (B), blue). In (A) the measurement time is $10^5$ and the  number of particles is $600$. In (B) the measurement time is $10^2$ and $6000$ realizations were used.  Panel (C): The PDF of the random variables $\xi$ which are defined in Eqs.~\eqref{eq:ZetaDefine} with ${\cal O}[x(t)]=x^2(t)$ [red stars, $t=10^3$, $10^5$ particles, $\sqrt{D(x)}=|\alpha|x^{1-1/\alpha}\theta(0\leq x\leq x_c)+\theta(x_c<x\leq L)$] and ${\cal O}[x(t)]=\theta(0.4<x(t)<0.6)$  [blue circles, $t=10^2$, $10^4$ particles, $\sqrt{D(x)}=x(1-x)$]. The black curve represents the analytic prediction Eq.~\eqref{eq:xi}.       }
\label{fig:mean_xi_x_1-x_}
\end{figure}

Furthermore, we consider the observable   ${\cal O}(x(t))=\theta (0.4<x(t)<0.6)$ which is integrable with respect to ${\cal I}_{\infty}(x)$. 
%Then we obtain
% \begin{eqnarray}
% \lim_{t\rightarrow\infty}\int_0^t {\rm d}t'\theta (0.4<x(t)<0.6) =  \nonumber \\
% = 2t^{1/2} \langle \theta (0.4<x<0.6) \rangle_{P_{\infty}} 
% &\approx&  0.647 t^{1/2}.
% \end{eqnarray}
% We define the random variable
% \begin{equation}
%  \xi   = \frac{\frac{1}{t}\int_0^t {\rm d}t'\theta (0.4<x(t)<0.6)}{  2\langle \theta (0.4<x<0.6) \rangle }
%  \label{eq:zeta2}
% \end{equation}
In Fig.~\ref{fig:mean_xi_x_1-x_} (blue circles) we show  that for the variable $\xi$, see Eq.~\eqref{eq:ZetaDefine}, we find $\langle \xi \rangle \rightarrow 1$ in the long time limit [panel (B)]. 
We show that ${\rm P}(\xi)$ is the Mittag-Leffler distribution of order $1/2$  Eq.~\eqref{eq:xi} [panel (C)], as expected, using same arguments as given in previous sections, so the Aaronson-Darling-Kac theorem applies.

Above, we study observables that are integrable with respect to the infinite density, so the PDF of these observables' time average is the Mittag-Leffler distribution. Here, since we have two non-integrable points, the sojourn times PDFs of the path $x(t)$ in $(0.5,1)$ and in $(0,0.5)$ are both power-law  $\psi(\tau)\sim\tau^{-3/2}$. In this case, the distribution of some other observables which are non-integrable with respect to ${\cal I}_{\infty}(x)$ are known as well. For example, in the long time limit, the occupation time in $(0.5,1)$ is distributed with the Lamperti distribution of order $1/2$ which corresponds to the arcsine law, see \cite{Godreche,Margolin06,akimoto2008generalized}.  

\section{Discussion}

Generally, the appearance of an infinite density with its peculiar non-integrable points is related to the classification of boundary points and to the interpretation of the Langevin equation  \cite{Gardiner,FellerBook,martin2011first}. The non-integrable point serve as a 'natural' boundary, which refers to a boundary which can neither be
reached in finite mean time nor be the starting point of a process. %A further discussion on the boundaries classification see  \cite{Gardiner,FellerBook,martin2011first}. 

For example, consider the process in $[0,L]$ when $\sqrt{D(x)}\sim x^{1-1/\alpha}$ where $x\rightarrow 0$ with $\alpha<0$ (e.g. see Sec.~\ref{sec:negative}) with Stratonovich interpretation. There, the heterogeneous diffusion coefficient is effectively slowing the particle in a sufficient way, so the divergent point $x\rightarrow 0$ actually serves as a 'natural' boundary since the particle never reaches the boundary, yet it approaches there slowly. Interestingly, $\sqrt{D(x)}\sim x^{1-1/\alpha}$ vanishes when $x\rightarrow 0$ for  $\alpha<0$ or $\alpha>1$, yet the infinite density appearance is related to the interpretation and the value of $\alpha$ itself. For It\^o and Stratonovich interpretations, we may find 
either an infinite density or a normalized one, within a finite domain, depends on $\alpha$. H\"anggi-Klimontovich interpretation is significantly different, since thermal equilibrium is attained for every process defined in a finite domain with a valid diffusivity (i.e. when the growth condition is fulfilled). This is so since, from its construction the H\"anggi-Klimontovich interpretation is built to yield a thermal state, which is a uniform distribution in a finite domain 
with reflecting boundary conditions and in the absence of external forces. 
 In Table \ref{Tab2} we present for different diffusivities the different regimes depend in $A$ and $\alpha$.     

For unbounded processes, where $\sqrt{D(x)}
 \sim x^{1-1/\alpha}$ for large $x$ and $\alpha>0$ (see, e.g. Secs.~\ref{sec:PurePowerLaw} and \ref{sec:Regularized}), the zero-current solution ${\cal I}_{\infty}(x)$ is non-integrable due its behavior at $x\rightarrow \infty$. In these cases $D(x)$ does not necessarily vanish anywhere, and there are no particular points where particles accumulate. Of-course, if the time-dependent distribution $P(x,t)$ becomes broader with time, the point $x\rightarrow\infty$ is a natural boundary. In Sec.~\ref{sec:Regularized} the effective force with the It\^o interpretation may limit the expansion of $P(x,t)$, so the equilibrium state is reached, see Tab~\ref{Tab2}.

% \subsection{Comment about the time-dependent diffusivity}
% We note that in our model the diffusion coefficient is time independent. Thus we find that the exponent corresponds to the time is given by $\alpha$ (the nature of the diffusion) and $A$ (the interpretation).  However, one may consider a  case where the diffusion coefficient depends on both space and time $\sqrt{D(x,t)}\propto x^{1-1/\alpha} t^{\gamma/2}$ where $-1<\gamma<1$, see e.g. \cite{cherstvy2015ergodicity}. Then one finds that for an observable ${\cal O}$ which is integrable with respect to the infinite density behaves as %$\langle \overline{\cal O} \rangle \rightarrow t^{(\gamma+\beta)/2}\langle {\cal O}\rangle _{P_{\infty}}/(\gamma+\beta)$ 
% $\langle \overline{\cal O} \rangle \rightarrow \langle {\cal O}\rangle /(\gamma+\beta)$
% in the long time limit, with $\beta$ defined in Eq.~\eqref{eq:beta}. In addition the PDF of the random variable %$\zeta \equiv \overline{\cal O}/[t^{\gamma/2+\beta/2}\langle O \rangle_{P_{\infty}}/(\gamma+\beta)]$ 
% $\zeta \equiv \overline{\cal O}/[\langle {\cal O} \rangle/(\gamma+\beta)]$ 
% should follows the Mittag-Leffler distribution of order $(\gamma+\beta)$. We leave the extension of the study to the tempo-spatial diffusion coefficient for a future research. 

\section{Summary}
We have shown that for processes with multiplicative
noise, in particular diffusion in inhomogeneous space,
infinite ergodic theory is the toolbox with which we analyze
the long-time behavior of the system.  
We have shown that the appearance of an infinite density is not related to the entire structure of $D(x)$ but it depends, together with the interpretation of the Langevin equation, on the behavior of $D(x)$ at large $x$ (for unbounded processes) or close to its zeros.
In particular we study processes with  $D(x)\sim |x-\tilde{x}|^{2-2/\alpha}$ in the vicinity of a point $\tilde{x}$. 
We examined the PDF $P(x,t)$ obtained from the Fokker-Planck equation corresponding to different interpretations of the Langevin equation;  It\^o, Stratonovich or H\"anggi-Klimontovich.
All these give rise to non-normalized densities, ${\cal I}_{\infty}(x)$, when the system is left unbounded, while It\^o and Stratonovich interpretations yield a nonnormalized state even for a bounded process. 
In the long-time limit we find $P(x,t) \rightarrow t^{\beta-1}{\cal I}_{\infty}(x)$, where $\beta$ is related to the first-passage-time distribution.

Furthermore, we consider observables ${\cal O}[x(t)]$ which are integrable with respect to the infinite density and show that the PDF of $\xi\equiv \lim_{t\rightarrow\infty}\beta {\overline{\cal O}}/\langle {\cal O}\rangle $ follows Mittag-Leffler distribution of order $\beta$ where $\langle \xi \rangle =1$.
This is in agreement with the Aaronson-Darling-Kac theorem.  One of the main results here is the identification of the relations between the exponents describing the diffusion field $D(x)$ and those  describing the nonnormalized state and the Mittag-Leffler statistics. For that we find useful two transformations Eqs.~\eqref{eq:LangevinMultiTranformed} and \eqref{eq:Transformation} 
which map the problem to Bessel processes (with purely logarithmic potential)  or regularized processes (where the potential is only asymptotically
logarithmic in some regime), so we can get finite time solutions. In particular, for the former we get exact solutions for all times, which is of benefit since it explains how the system approaches the nonnormalized state and in what sense. 

We note that all along our work we considered the time averages, which start at the moment of initiation of the process. That in a diffusion process corresponds to a medium, in which a particle is inserted at some time which we call the origin of time. However, we may choose to start measuring  at some time $t_a$, for example perform a time average in a window $(t_a, t_a + t)$ and this would give aging effects.
In deterministic setting the modification of the Aaronson-Darling-Kac theorem was considered in \cite{akimoto2013aging}, and it might be worthy to consider this more general scenario in the context of theory of multiplicative processes.

\begin{widetext}

\begin{table}
\centering
\begin{tabular}{|ccccp{5cm}|}
\hline \hline
 \multirow{2}{*}{$\sqrt{D(x)}$ } & bounded/  &  \multirow{2}{*}{Sec.} & & \\
  & unbounded  &    & &
\\ \hline \hline
\multirow{3}{*}{$\sqrt{D(x)}\propto x^{1-1/\alpha}$ }  & \multirow{3}{*}{unbounded} & \multirow{3}{*}{\ref{sec:PurePowerLaw}} & & Region I - ${\cal I}_{\infty}(x)$ non-integrable, the non integrable point is $x\rightarrow \infty$ \\
  &  &  & & Region II - $P(x,t)$ non-exist \\ 
  & & &
\includegraphics[trim= 90 250 90 380, width=0.2\columnwidth]{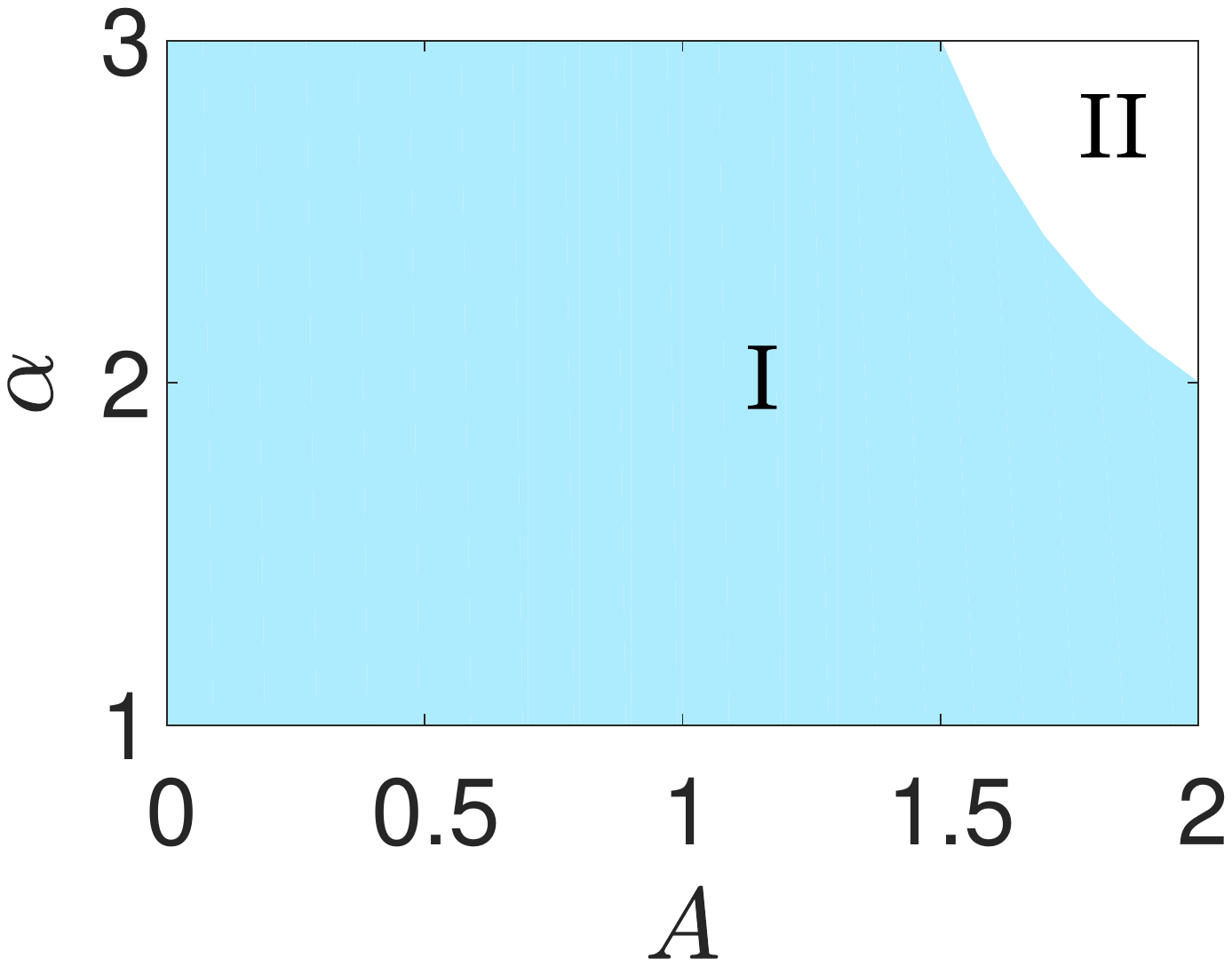} & 
\\
\hline
\multirow{3}{*}{$\sqrt{D(x)}\propto x^{1-1/\alpha}\theta(x\geq x_c)+\theta(x<x_c)$}  & \multirow{3}{*}{unbounded} & \multirow{3}{*}{\ref{sec:Regularized}} & & Region I - ${\cal I}_{\infty}(x)$ non-integrable, the non integrable point is $x\rightarrow \infty$  \\
  &  &  & & Region II - ergodic phase \\ 
  & & &
\includegraphics[trim= 90 250 90 380, width=0.2\columnwidth]{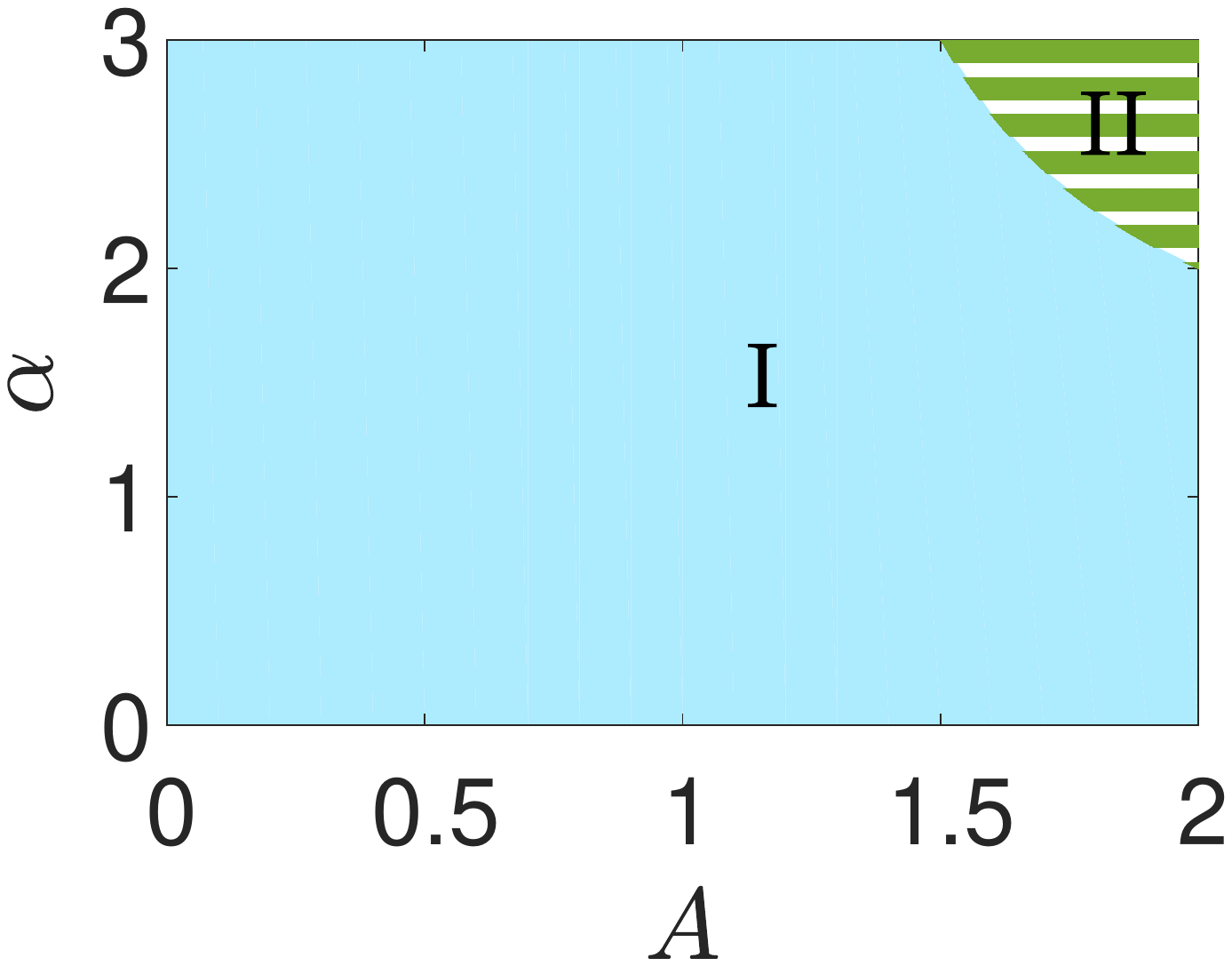} & 
\\
 \hline
$\sqrt{D(x)}\propto x^{1-1/\alpha}$ & \multirow{2}{*}{bounded} & \multirow{2}{*}{ \ref{sec:negative}} & & Region I - ergodic phase\\ 
  $\sqrt{D(x)}\propto x^{1-1/\alpha}\theta(x_c<x<L)+\theta(x>x_c)$ &  &  & &  Region II - ${\cal I}_{\infty}(x)$ non-integrable, the non integrable point is $x\rightarrow 0$ Region III - Growth condition is not fulfilled - processes are not defined \\ 
    & & &
\includegraphics[trim= 90 250 90 470, width=0.2\columnwidth]{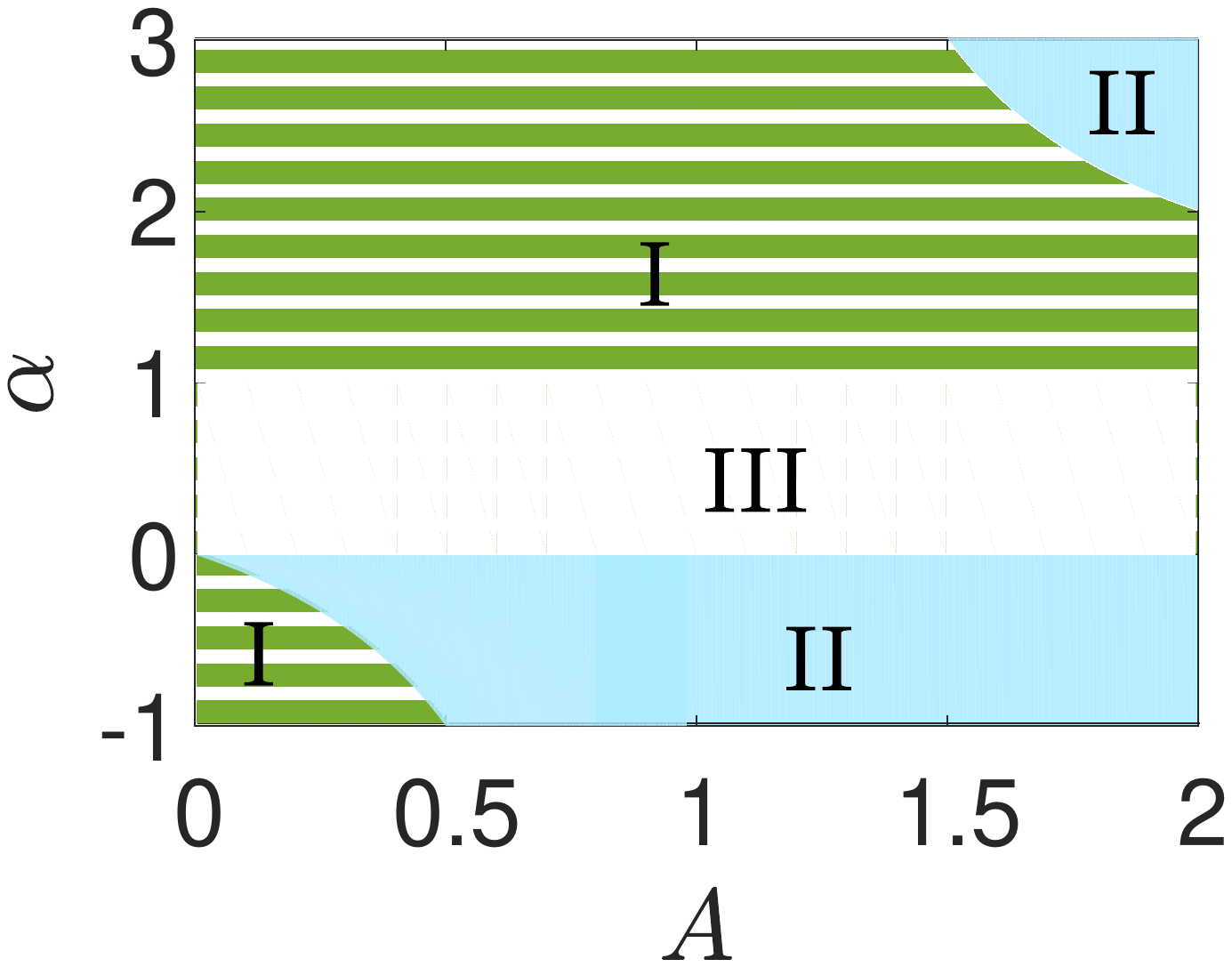} & 
\\
\hline \hline
\end{tabular}
\caption{The different regimes, infinite density and ergodic phases, depend on $\alpha$ and $A$ for some diffusivities examine in the paper. Here $A$ is considered a continuous parameter, though as was mentioned usually $A=0,1,2$ for I, S, and HK interpretations respectively. Only for convenience we limit the plots to $\alpha<3$, though $\alpha$ can attains values greater than $3$. In the upper and the middle plots for values $\alpha<1$ and $\alpha<0$ (respectively) a solution $P(x,t)$ does not exist. In the lower panel, there is no lower or upper bounds to $\alpha$ and the choice of presenting $-1<\alpha<3$ is arbitrary.    }
\label{Tab2}
\end{table}

\end{widetext}

\acknowledgements
The support of Israel Science Foundation's grant 1898/17 is acknowledged.
We thank Guenter Radons and Takuma Akimoto for the discussion and comments.

\appendix

\section{Growth Condition}
\label{App:Growth}
A mathematical issue arises when considering the process describe by Langevin equation \eqref{eq:LangevinMulti} as is hereby explained. The following conditions guarantee the existence and uniqueness of the solution of the Langevin equation Eq.~\eqref{eq:LangevinMulti}: a $K \in {\mathbb{R^+}}$ exist such that for every time \cite{Gardiner}
\begin{eqnarray}
\left|\sqrt{D(x)}-\sqrt{D(y)}\right|\leq K |x-y| & {\rm (Lipschitz\ condition)}, \nonumber\\
D(x) \leq K^2 (1+x^2) & {\rm (growth\ condition)}. \nonumber
\end{eqnarray}
These conditions should be satisfied for $x$ and $y$ in the underlined interval of a given process.  Therefore the following requirements are taken:

% \begin{table}[h!]
\begin{center}
{ \centering
		\begin{tabular}{cccccccc}
	\hline \hline 
   $\sqrt{D(x)}$ & Section & $\alpha$ \\ 
   \hline 
   pure power-law in $[0,\infty)$  & \ref{sec:PurePowerLaw} & $\alpha\geq 1$ \\
   regularized process in $(-\infty,\infty)$ & \ref{sec:Regularized}& $\alpha>0$ \\
   power-law behavior close to zero in $[0,L]$ & \ref{sec:negative} & $\alpha<0$ \\
     &  & $\alpha\geq 1$ \\
   \hline \hline
    \end{tabular}
    }
\end{center}
%\end{table}

We note though that violation of the conditions does not necessarily mean that there is no solution, it rather means that the solution of Eq.~\eqref{eq:LangevinMulti} might diverge at finite time and thus does not describe a physical behavior. See further discussion in \cite{Gardiner}.   

\section{Stratonovich Interpretation with external force approaches H\"anggi-Klimontovich and It\^o forms   }
\label{Sec:Appendix1}
Consider the following Fokker-Planck equation
\begin{equation}
\frac{\partial P(x,t)}{\partial t} = \frac{1}{2}\frac{\partial}{\partial x}\left[D(x) \frac{\partial}{\partial x} P(x,t)\right],
\end{equation}
which is related to the Langevin equation \eqref{eq:LangevinMulti} with H\"anggi-Klimontovich interpretation. This equation may also be written as 
\begin{eqnarray}
\frac{\partial P(x,t)}{\partial t} &=& \frac{1}{2}\frac{\partial}{\partial x}\left[D(x) \frac{\partial}{\partial x} P(x,t)\right]= \nonumber \\
&=& \frac{1}{2}\frac{\partial}{\partial x}\left[\sqrt{D(x)} \frac{\partial}{\partial x} \sqrt{D(x)}(x,t)\right] \nonumber \\
&-&\frac{1}{2}\frac{\partial }{\partial x}\left[\sqrt{D(x)}\frac{\partial \sqrt{D(x)}}{\partial x}P(x,t)\right].
\end{eqnarray}
Thus,  the Stratonovich interpretation of a Langevin equation with an external potential is equivalent to the Langevin equation \eqref{eq:LangevinMulti} with the H\"anggi-Klimantovich approach. Its corresponding Langevin equation is
\begin{equation}
\frac{dx}{dt}=\sqrt{D(x)} \eta(t) + \frac{1}{2}\sqrt{D(x)}\frac{d\sqrt{D(x)}}{dx}.
\end{equation}
Using the transformation $y(x)=\int (D(x))^{-1/2} dx$ (which, as mentioned, may be applied just on Stratonovich interpretation)  we find
 \begin{equation}
\frac{dy}{dt}= \eta(t) + \frac{1}{2}\frac{d\sqrt{D(y)}}{dy}\frac{1}{\sqrt{D(y)}}.
\end{equation}
The additional effective force, which is proportional to $ {d_y \sqrt{D(y)}}/{\sqrt{D(y)}}$, is determined by the properties of $\sqrt{D(y)}$. Using Eq.~\eqref{eq:Diff} and we obtain
\begin{equation}
\frac{dy}{dt}=\eta(t)-\frac{(1-\alpha)/2}{y}
\end{equation}
which is a Brownian motion in a logarithmic potential as given in Eq.~\eqref{eq:y} with $A=0$. 

In a similar fashion the Fokker-Planck equation corresponding It\^o interpretation may be presented as
\begin{eqnarray}
\frac{\partial P(x,t)}{\partial t} &=& \frac{1}{2}\frac{\partial^2}{\partial x^2}\left[ D(x) P(x,t)\right]= \nonumber \\
&=& \frac{1}{2}\frac{\partial}{\partial x}\left[\sqrt{D(x)} \frac{\partial}{\partial x} \sqrt{D(x)}(x,t)\right] \nonumber \\
&+&\frac{1}{2}\frac{\partial}{\partial x}\left[\sqrt{D(x)}\frac{\partial \sqrt{D(x)}}{\partial x}P(x,t)\right],
\end{eqnarray}
therefore 
\begin{equation}
\frac{dy}{dt}=\eta(t)+\frac{(1-\alpha)/2}{y},
\end{equation}
hence we recover Eq.~\eqref{eq:y} with $A=2$.

\section{First Passage Time when $-1<U_0<1$}
\label{AppC}
Following \cite{bray2000random}, the distribution of $y$ at time $t$ with absorbing boundary at $y=0$ is 
\begin{equation}
\tilde{P}_{\rm abs}(y,t;y_0,0)=e^{-\frac{y^2+y_0^2}{2t}}y_0^{\frac{1}{2}+\frac{U_0}{2}}y^{\frac{1}{2}-\frac{U_0}{2}}I_{\frac{1}{2}+\frac{U_0}{2}}\left(\frac{y_0y}{t}\right)\frac{1}{t}.
\end{equation}
This solution exists (i.e normalizable) when $U_0>-1$.
The current thus trough the origin $y=0$ is
\begin{eqnarray}
J(0)&\equiv& -\frac{1}{2}\left[\frac{\partial \tilde{P}_{\rm abs}(y,t)}{\partial y}+\frac{U_0}{y}\tilde{P}_{\rm abs}(y,t)\right]_{y=0} 
\\ \nonumber &=& \frac{2^{-\frac{U_0}{2}-\frac{1}{2}}
   t^{-\frac{U_0}{2}-\frac{3}{2}} \exp\left(-\frac{y_0^2}{2 t}\right)
   y_0^{U_0+1}}{\Gamma
   \left[\frac{U_0}{2}+\frac{1}{2}\right]}
   \stackrel{t\rightarrow\infty}{\propto} t^{-\frac{U_0}{2}-\frac{3}{2}} 
\end{eqnarray}
which is a known result, see e.g. \cite{bray2000random}.
Back from $y$ to $x$ we find that the current of the probability at $x=0$ behaves the same, i.e. $\propto t^{-(U_0+3)/2}$. In a similar fashion one can prove that the probability of sojourn times $\tau$ outside a finite subspace $(x_1,x_2)$ follows
\begin{equation}
\psi(\tau) \sim \tau^{-1-\beta} \ \ \ \ \ {\rm with \ \ \ \ \  } \beta=\frac{1+U_0}{2}.
\end{equation}
Using renewal processes theory \citep{Godreche} we find that $\xi$ defined in Eq.~\eqref{eq:ZetaDefine} is distributed via PDF Eq.~\eqref{eq:zeta} as given in the maim text.

% \subsection{Strong Repulsive Potential: $U_0\leq -1$}
% There is no solution for Eq.~\eqref{eq:FP:General} with diffusion coefficient Eq.~\eqref{eq:Diff} with absorbing boundary at $x=0$. That means that there is no recurrence and particle never return back to zero. 

% \subsection{Strong Attractive Potential: $U_0\geq 1$}
% In this case, when diffusion close to zero is normal as is discussed in Sec.~\ref{sec:Regularized}, there exists a normalizable steady state and the process is ergodic. 

\section{Derivation of the distribution of  $\xi$}
\label{AppD}
Consider a renewal process where the events occur at the random epoch, and the waiting times between the events distributed with 
\begin{equation}
\psi(\tau)\sim \tau^{-1-\beta}
\label{eq:tauDist}
\end{equation}
in the long time limit. The PDF of $n$ renewals up to time $t$ in Laplace space is
\begin{equation}
{\cal L}\left\{{\rm PDF}[n]\right\}=\psi^n(s)\frac{1-\psi(s)}{s}.
\label{eq:LaplacePDF}
\end{equation}
In the following we calculate this PDF for several cases of $\psi(s)$.

\subsubsection*{Case 1: Infinite mean sojourn time, i.e. $0<\beta<1$}

In this case, in the small $s$ limit the the Laplace transform of Eq.~\eqref{eq:tauDist} reads
\begin{equation}
\psi(s)\sim 1-\Gamma(1-\beta)s^{\beta}+ \ldots 
\label{eq:D3}
\end{equation}
By substituting into Eq.~\eqref{eq:LaplacePDF} we find
\begin{equation}
{\cal L}\left\{{\rm PDF}[n]\right\}\approx e^{-\Gamma(1-\beta)n s^{\beta}} \Gamma(1-\beta)s^{\beta-1},
\label{eq:D4}
\end{equation}
where the small $s$ limit is taken. Its inverse Laplace transform gives
\begin{equation}
{\rm PDF}[n]\approx \frac{t}{\beta n ^{1+\frac{1}{\beta}}\Gamma(1-\beta)^{\frac{1}{\beta}}}L_{\beta}\left[\frac{t}{\Gamma(1-\beta)^{\frac{1}{\beta}}n^{\frac{1}{\beta}}}\right].
\end{equation}
The mean of $n$ is given by
\begin{eqnarray}
&&{\cal L}\left[\langle n \rangle \right] = \frac{\psi(s)}{s[1-\psi(s)]} \approx \frac{s^{-\beta-1}} {\Gamma(1-\beta)} \\
&& \Rightarrow \langle  n (t) \rangle \approx \frac{t^{\beta}}{\Gamma(1-\beta)\Gamma(1+\beta)}. 
\nonumber
\end{eqnarray}
Now, we define the scaling variable
\begin{equation}
\xi\equiv  \frac{n}{\langle n \rangle } =  \Gamma(1-\beta)\Gamma(1+\beta) \frac{n}{t^{\beta}}.
\end{equation}
Therefore, the distribution of $\xi$ is 
\begin{equation}
{\rm P}(\xi)={\cal M}_{\beta}(\xi)\equiv \frac{\Gamma ^{\frac{1}{\beta}}(1+\beta)}{\beta \xi ^{1+\frac{1}{\beta}}} L_{\beta}\left[\frac{\Gamma ^{\frac{1}{\beta}}(1+\beta)}{\beta \xi ^{\frac{1}{\beta}}}\right],
\end{equation}
as is give in Eq.~\eqref{eq:zeta}.

\subsubsection*{Case 2: Finite mean sojourn time}

In this case, in the small $s$ limit the the Laplace transform of Eq.~\eqref{eq:tauDist} reads
\begin{equation}
\psi(s)\sim 1-\langle \tau\rangle  s+ ...
\end{equation}
By substituting into Eq.~\eqref{eq:LaplacePDF} we find
\begin{eqnarray}
&&{\cal L}\left\{{\rm PDF}[n]\right\}\approx e^{-n \langle \tau\rangle  s } \langle \tau \rangle ,
\\
&&\rightarrow {\rm PDF}[n]\approx \langle \tau \rangle \delta\left(t-n\langle \tau \rangle \right),
\nonumber
\end{eqnarray}
in the long time (small $s$) limit. We define 
\begin{equation}
\xi \equiv \frac{n}{\langle n \rangle} = \frac{n \langle \tau \rangle }{t},
\end{equation}
so 
\begin{equation}
{\rm P}(\xi)= \delta (\xi-1).
\label{eq:D12}
\end{equation}
In Sec.~\ref{sec:Regularized} with It\^o interpretation, a case where $\beta > 1$ is possible, so the sojourn times' PDF has a mean. In these cases Eq.~\eqref{eq:D12} is obtained as is given in Eq.~\eqref{eq:Alpha2} for $\beta \rightarrow 1$ (equivalent to $\alpha \rightarrow 2$).

\subsubsection*{Case 3: Limit of $\beta \rightarrow 0$}

 This case where $\beta\rightarrow 0$, is delicate since we should take both small $s$ limit and the limit when $\beta$ approaching to zero from above.  
 %Generally, to evaluate a PDF one can use its expansion to moments.
% \begin{equation}
% {\cal L}\left\{{\rm PDF}[n]\right\} = 1 + \langle n \rangle s + \frac{\langle n ^2 \rangle }{2!}s^2 +  \frac{\langle n ^3 \rangle }{3!}s^3 + \ldots .  
% \end{equation}
Importantly, we first take the limit of small $s$ and $0<\beta<1$, and then we calculate the limit when $\beta \rightarrow 0$. Using Eq.~\eqref{eq:D4} we obtain
% \begin{eqnarray}
% &&{\cal L}\left\{{\rm PDF}[n]\right\}\approx e^{-\Gamma(1-\beta)n s^{\beta}} \Gamma(1-\beta)s^{\beta-1} \nonumber \\ \nonumber
% && = \frac{\Gamma(1-\beta)}{s^{1-\beta}}\left[1-\Gamma(1-\beta)n s^{\beta}+ \frac{\Gamma^2(1-\beta)n^2}{2!} s^{2\beta} + \ldots 
% \right. \\
% && \left. \ldots + \frac{\Gamma^k(1-\beta)n^k}{k!} s^{\beta k} + \right] \rightarrow \frac{1}{s}\left[1-n+\frac{n^2}{2}+ \ldots 
% \frac{n^k}{k!} + \right] \nonumber \\
% && \frac{1}{s} \exp(-n)
% \end{eqnarray}
\begin{eqnarray}
{\cal L}\left\{{\rm PDF}[n]\right\}\approx e^{-\Gamma(1-\beta)n s^{\beta}} \Gamma(1-\beta)s^{\beta-1}   \\
\Rightarrow {\cal L}\left\{{\rm PDF}[n]\right\} \stackrel{\beta \rightarrow 0}{\longrightarrow}   \frac{1}{s} \exp(-n) \nonumber
\end{eqnarray}
and its inverse Laplace transform gives
% \begin{eqnarray}
% &&{\cal L}\left[\langle n \rangle \right]=\frac{s^{-1-\beta}}{\Gamma(1-\beta)} \stackrel{\beta \rightarrow 0}{\rightarrow} \frac{1}{s} \Rightarrow \langle n(t) \rangle \approx 1 
% \nonumber
% \\
% &&{\cal L}\left[\langle n^2 \rangle \right]=\frac{2 s^{-1-2\beta}}{\Gamma^2(1-\beta)} \stackrel{\beta \rightarrow 0}{\rightarrow} \frac{2}{s} \Rightarrow \langle n^2(t) \rangle \approx 2
% \nonumber \\
% &&{\cal L}\left[\langle n^3 \rangle \right]=\frac{6s^{-1-3\beta}}{\Gamma(1-\beta)} \stackrel{\beta \rightarrow 0}{\rightarrow} \frac{6}{s} \Rightarrow \langle n^3(t) \rangle \approx 6
% \nonumber
% \\
% && \ \ \ \ \ \ \ \ \ \ \ \ \ \ \ \ \ \ldots \ \ \ \ldots  \ \ \ \ \ \ \ \ \Rightarrow \langle n^k (t) \rangle \approx k! 
% \nonumber
% \\
% \ 
% \end{eqnarray}
% Therefore, we obtain
% \begin{equation}
% {\cal L}\left\{{\rm PDF}[n]\right\} = 1 +  s + s^2 + s^3 + \ldots  = \frac{1}{1+s} 
% \end{equation}
% hence
% \begin{equation}
% {\rm PDF}[n]= {\cal L}^{-1}\left[\frac{1}{1+s}\right] = \exp(-n).
% \end{equation}
\begin{equation}
{\rm PDF}[n]= {\cal L}^{-1}\left[\frac{1}{s}\exp(-n)\right] = \exp(-n).
\end{equation}
The rescaled variable is $\xi \equiv n/\langle n \rangle =n$ (since $\langle n\rangle =1$) so 
\begin{equation}
{\rm P}(\xi)= \exp(-\xi)
\end{equation}
as is given in Eq.~\eqref{eq:Alpha2} in the main text.

\section{Simulation Methods}
There are mainly two methods to simulate Langevin equation with multiplicative noise regard to the different interpretations. 

\subsubsection*{ Method 1:}

This method is based on the fact that for variable $y$ [given from the mapping $y(x)$] the process evolves with additive noise instead of multiplicative noise. The algorithm is as follows
\begin{enumerate}
\item Transforming to variable $y(t)\equiv y[x(t)]$ as given for example in Eq.~\eqref{eq:y} in
the main text.
\item Using Euler discretization
\begin{equation}
y(t+\delta t)= y(y)+\eta\delta t - \frac{U_0/2}{y}\delta t.
\end{equation}
\item Transforming back following $x(t+\delta t)=y^{-1}(t+\delta t)$.
\end{enumerate} 

\subsubsection*{Method 2:}

Here, one can use the known transformations between the interpretations: It\^o $\leftrightarrow$ Stratonovich $\leftrightarrow$ H\"anggi-Klimontovich with addition of effective force. Since Euler discretization may be applied on It\^o interpretation solely, we used the transformation from other interpretations to It\^o and find
\begin{equation}
\dot{x}(t)=\sqrt{D(x)}\eta(t)+\frac{2-A}{2}\sqrt{D(x)}\frac{\partial \sqrt{D(x)}}{\partial x}
\end{equation}
which is now interpret via It\'o and we may use Euler discretization, i.e.
\begin{equation}
x(t+\delta t)=x(t)+\sqrt{D[x(t)]}\eta \delta t+\frac{2-A}{2}\sqrt{D(x)}\frac{\partial \sqrt{D(x)}}{\partial x}\delta t.
\end{equation}
For both methods one should use small $\delta t$.

\bibliography{./bib1}
\end{document}